\documentclass[namedreferences]{solarphysics}
\usepackage[optionalrh,showbiblabels]{spr-sola-addons} 
\usepackage{graphicx}        
\usepackage{color}           
\usepackage{breakurl}        
\usepackage{pdflscape}
\usepackage{ulem}
\usepackage{gensymb}
\usepackage{enumitem}

\begin{document}

\begin{article}
\include{defs}
\begin{opening}

\title{The LASCO Coronal Brightness Index}

\author[addressref={aff1},corref,email={karl.battams@nrl.navy.mil}]{\inits{K.}\fnm{Karl}~\lnm{Battams}\orcid{0000-0002-8692-6925}}
\author[addressref={aff1},email={russell.howard@nrl.navy.mil}]{\inits{R.A.}\fnm{Russell~A.}~\lnm{Howard}\orcid{0000-0001-9027-8249}}
\author[addressref={aff1},email={hillary.dennison@nrl.navy.mil}]{\inits{H.}\fnm{Hillary~A.}~\lnm{Dennison}\orcid{0000-0002-1582-5027}}
\author[addressref={aff2},email={rweigel@gmu.edu}]{\inits{R. S.}\fnm{Robert~S.}~\lnm{Weigel}\orcid{0000-0002-9521-5228}}
\author[addressref={aff1},email={judith.lean@nrl.navy.nil}]{\inits{J.}\fnm{Judith~L.}~\lnm{Lean}\orcid{0000-0002-0087-9639}}

\address[id=aff1]{US Naval Research Laboratory, Washington D.C., USA}
\address[id=aff2]{George Mason University, Fairfax VA, USA}
%
\runningauthor{Battams et al.}
\runningtitle{LASCO Coronal Brightness Index}

\begin{abstract}
We present the construction of a new white-light coronal brightness index (CBI) from the entire archive of observations recorded by the \textit{Large Angle Spectrometric Coronagraph} (LASCO) C2 camera between 1996 and 2017, comprising two full solar cycles. We reduce all fully calibrated daily C2 observations of the white light corona into a single daily coronal brightness observation for every day of observation recorded by the instrument, with mean daily brightness values binned into 0.1$\mathrm{R}_\odot$ radial $\times$ 1 degree angular regions from 2.4 -- 6.2$\mathrm{R}_\odot$ for a full 360-degrees. As a demonstration of the utility of the CBI, we construct a new solar irradiance proxy that correlates well with a variety of direct solar irradiance observations, with correlations shown to be in the range of 0.77--0.89. We also present a correlation mapping technique to show how irradiance correlations depend on, and relate to, coronal structure/locations, and to demonstrate how the LASCO CBI can be used to perform long-term ``spatial correlation'' studies to investigate relationships between the solar corona and any arbitrary concurrent geophysical index. Using this technique we find possible relationships between coronal brightness and plasma temperature, interplanetary magnetic field magnitude and (very weakly) proton density.
\end{abstract}

%

\end{opening}

%

\section{Introduction} \label{s:intro}

The \textit{Large Angle Spectrometric Coronagraph} (LASCO) telescope aboard the joint ESA/NASA \textit{Solar and Heliospheric Observatory} (SOHO) mission has been operating almost continuously for over twenty-three years, returning white-light observations of the solar corona from 2.5--30 $\mathrm{R}_\odot$. The continuity and stability of this instrument, coupled with its precise photometric calibrations, present the opportunity for the investigation of coronal intensities from a single instrument over a multi-solar cycle timeline.

Key features of interest in the LASCO observations are often coronal mass ejections (CMEs) and their properties \citep[\textit{e.g.}][]{StCyr00,Vourlidas10,Webb12}, but the data also hold valuable information regarding other coronal features such as solar streamers \citep[\textit{e.g.}][]{Wang97,Thernisien06,Feng11}, and both the F (dust) and K (electron) corona \citep[\textit{e.g.}][]{Andrews03,Morgan07, Shopov08}. With more than two complete solar cycles of observations from LASCO, we now have the ability to examine these and other features in the much broader context of the solar cycle. In particular we can now investigate the global brightness of the solar corona, and by proxy its electron content, over this long time period. Early related studies, such as \cite{Andrews03}, looked at the variation of coronal brightness in relation to the solar cycle in the early LASCO mission, noting both an increase in brightness and complexity of coronal structures as the cycle progressed towards maximum. In 2012, an examination of coronal brightness extracted from radial scans of LASCO C2 observations were first presented by \cite{Dennison12}, who also demonstrated that the resulting time series correlated well with various solar irradiance measures. This technique was used by \cite{Lamy14} to examine LASCO C2 observations of the inter-minima brightness of solar cycles 22/23 and 23/24, and then furthered by \cite{Barlyaeva15} who again demonstrated the correlation of LASCO coronal brightness to solar irradiance and other possible geophysical indices using a technique that extracted observations from radial scans and latitudinal sectors in the C2 field of view.

The Sun is the driver of all space weather conditions at Earth, indirectly and directly driving a number of important geophysical indices, such as solar irradiance \citep{Willson1978, Kopp11}, 10.7cm solar radio flux \citep{Greenkorn2012, Tapping13}, proton temperature and density \citep{Neugebauer66,Elliott16}, interplanetary magnetic field (IMF) strength \citep{Chang73,Boyle97}, among several others. Thus a secondary benefit to the longevity and stability of the LASCO observations is that they can be used to study -- and in some cases provide proxies of -- some of these indices. For example, the Sun provides a total irradiance of 1360.8 Wm$^{-2}$ at the average Sun-Earth distance of 1 Astronomical Unit (AU, \cite{Kopp11, Schmutz2013, Pra2016}). Changes in solar irradiance may alter Earth's climate, atmosphere and ionosphere through radiative, chemical dynamical processes \citep{Gray10, Lean16}. Reliable knowledge of solar irradiance variability on time scales of the 11-year solar cycle and longer is thus crucial for understanding how and why Earth's surface, atmosphere and ionosphere vary in comparison with other natural and anthropogenic influences.

Some prior work suggests that total solar irradiance (TSI) was lower by more than 0.2Wm$^{-2}$ in the prolonged solar minimum epoch of 2008-2009, at the end of solar cycle 23, than at the beginning of the cycle in 1996 \citep{Frohlich09, Russell10}. This purported inter-minima decrease in total irradiance, which occurred even as Earth's global surface temperature increased 0.2${\degree}$C, has been cited as evidence that solar irradiance is not a significant cause of changes in Earth's surface temperature in recent decades \citep{Hansen11}. \cite{Frohlich09} find no accompanying decrease in chromospheric indices and UV irradiance, and hypothesize that the cause of the TSI changes is a long-term change in the Sun's global temperature that does not influence the UV emissions from the solar atmosphere. In contrast, other work suggests that solar EUV irradiance did in fact decrease by 15\% from the 1996 to 2008-2009 solar cycle minima \citep{Didkovsky10}. Resolving the causes of decadal trends in geophysical parameters, whether at Earth's surface or in space, requires reliable specification of those indices over time scales of the solar cycle and longer. The reality (or not) of inter-minima solar irradiance changes is controversial because of the relative shortness of extant irradiance databases and their (generally) inadequate long-term repeatabilities, which result in uncertainties that exceed the magnitude of true solar changes. Since the individual observations typically do not extend for multiple solar cycles, overlap and cross calibration is crucial to establish reliable long-term irradiance records. Recently, \cite{Kopp18} have proposed a method to combine the separate TSI data sets into a long-term data set, but it should be noted that producing high-resolution TSI measurements is extremely challenging. As we demonstrate in this article, the white-light intensity LASCO observations can be used to derive stable, long-term proxies of certain solar indices, including TSI. This not only provides a potential calibration baseline for other measurements, but also enables investigations of the spatial/structural relationships between the solar K corona and these various indices.

We present here the completion of the methodology first described by \cite{Dennison12}, to reduce the entire LASCO C2 mission catalog into a so-called Coronal Brightness Index (CBI) -- a compact, three-dimensional data set comprising spatially-located coronal brightness indices extracted at 0.1{$\mathrm{R}_\odot$}$\times$360-degree ($r\theta$) locations within the LASCO C2 field of view at a 1-day cadence for the mission duration. All raw instrument observations are fully calibrated using the most up-to-date calibration data available from the instrument team, with extracted data precisely spatially co-aligned and corrected for various operational effects. Following a detailed presentation of the steps taken to produce the CBI, and information regarding data sharing/availability of the index, we present two examples highlighting the utility of CBI. In our first example, we expand upon and complete our early work \citep{Dennison12} of producing several LASCO-based total and spectral solar irradiance indices. Our second example demonstrates the use of a ``correlation mapping'' technique that exploits the three-dimensional nature of the CBI to explore spatial relationships between the global coronal structure and any arbitrary geophysical index.

\section{LASCO C2 Coronal Observations}
The Large Angle and Spectrometric Coronagraph (LASCO) \citep{brueckner95} was launched 6 December 1995 on the Solar and Heliospheric Observatory (SOHO) \citep{Domingo95} into a halo orbit about the L1 Lagrangian point (in the sun-earth line, a distance 1.5 million km from the Earth).

\begin{figure*}
  \centering
  \includegraphics[width=110mm]{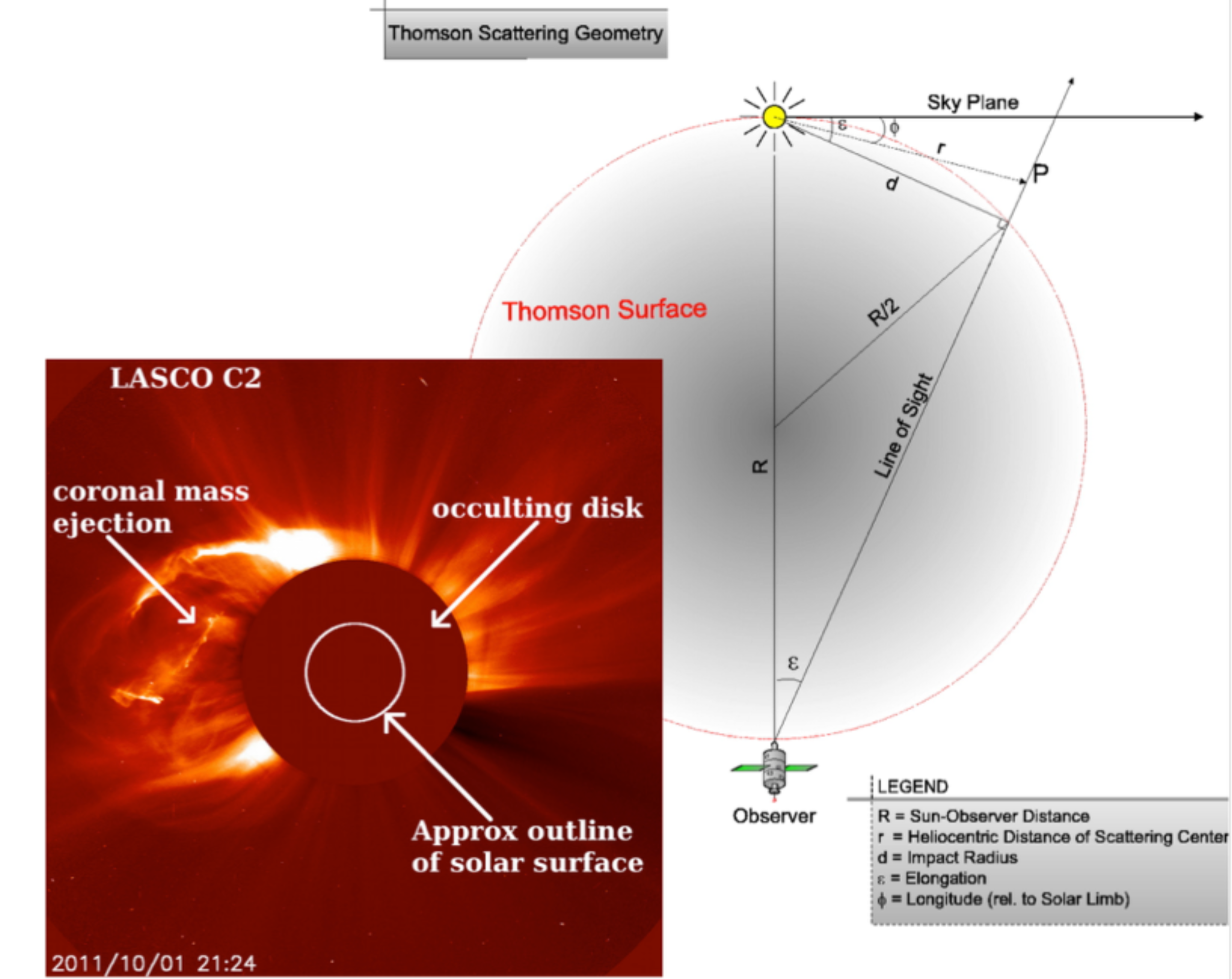}
  \caption[Images]{Shown is a schematic of the LASCO observing geometry for measuring white light (Thomson) scattered by the Sun's corona. Reduction of the LASCO detector signals of the spatial distribution of the Thomson scattered signal provides images of the white light solar corona, illustrated by the image. Indicated in the image are the central occulting disk and the solar limb. A coronal mass ejection (CME) is in progress on the left-hand side of the image, and coronal streamers can be seen pointing radially from the Sun in most locations.}
\label{fig:scattering_geom}
\end{figure*}

LASCO observes visible radiation that electrons in the Sun's corona from 1.1 to 30{$\mathrm{R}_\odot$} scatter into the instrument's field of view, as the schematic in Figure~\ref{fig:scattering_geom} depicts. This Thomson scattering, details of which were established for the solar corona in the 1950s and 60s \citep{Billings}, occurs preferentially on the Thomson surface \citep{Vourlidas06}. A typical LASCO image, which Figure~\ref{fig:scattering_geom} also illustrates, consists of four different components, namely the Thomson scattering of photospheric light from free electrons (the K-corona), Mie-like scattering of photospheric light from dust particles in orbit about the sun (the F-corona), light from planets, stars and other galactic sources and instrumental stray light. Reduction of the detector signals to brightness quantities taking into account the viewing geometry and separation of the components is well established \citep{Vourlidas06, Morrill06, Vourlidas10}.

Because the coronal visible intensity falls off very rapidly with height above the Sun's surface (Figure~\ref{fig:scattering_geom}), the LASCO instrument utilizes three telescopes to observe the entire range, each optimized for narrower height regimes with overlap of adjacent height regions providing inter-calibration.  Coronagraphs designated C1, C2 and C3 observe the coronal white light signal from 1.1 - 3$\mathrm{R}_\odot$, 2.5 - 6$\mathrm{R}_\odot$ and 3.7 - 30$\mathrm{R}_\odot$, respectively. For constructing a coronal white light irradiance index, we use observations made by the fully calibrated (level-1) LASCO C2 data recorded by a 1024$\times$1024 pixel CCD camera detector, which has a spatial resolution of 12 arc-seconds/pixel. We used the nominal full-resolution data product, which utilizes an orange filter with a wavelength transmittance of 540-640nm.

LASCO's routine observations of the solar corona began in May 1996, following an extended period of positioning into the halo orbit and commissioning, and continue to the present.  The most significant interruption in observations occurred in June 1998 when the spacecraft lost Sun pointing \citep{Fleck}. Also, at this loss of sun-pointing, the LASCO instrument became very cold (about -80C), and several components failed, most notably the C1 telescope. Intermittent interruptions followed until the spring of 1999 when routine observations resumed.  Then, in July 2003, as a result of a failure in the vertical drive of the high-gain antenna, spacecraft operations were reprogrammed to roll the spacecraft $180^{o}$ to point the antenna to Earth. These maneuvers occur every three months and can result in data losses of up to three days, but most typically of only a few hours.

\subsection{LASCO Data Preparation}

Securing a reliable data set of solar brightness variations from LASCO observations with the repeatability needed to investigate real long-term changes in the corona requires quantitative determination of all factors that affect the instrument's long-term stability. As part of the calibration procedure \citep{Morrill06}, a statistical measure of the photometric stability is obtained over a 28-day period.  The effect of the stability is that a movie of adjacent images over any time scale shows no flicker.  This stability (or relative precision) is 0.01\% for C2 and 0.1\% for C3.  The absolute photometric accuracy of the LASCO coronal brightness has been determined by measuring the instrument response to stars transiting the field of view.  Fitting the yearly transits of up to 500 stars produced standard deviations between 0.02 and 0.05 \citep{Colaninno}.  a constant degradation was observed to be 0.2\% annually \citep{Gardes,Colaninno} due to darkening of the optics from energetic particle radiation.  This constant degradation rate is included in the calibration of the instrument for the entire operational period of LASCO. This results in a uniform response for the entire period.  No degradation in the stability has been observed.

The LASCO CBI is derived from the ``Level-1'' data product, produced and hosted by the instrument principal investigators at the Naval Research Laboratory following procedures and calibration coefficients outlined in \cite{Gardes}, \cite{Colaninno} and \cite{Morrill06}. In short, this process produces the final fully-calibrated LASCO product in which optical distortions and vignetting are corrected, data units are converted to photometrically calibrated physical units of mean solar brightness (MSB), data file (FITS) headers are populated with finalized metadata, and observation and exposure times are corrected for the known failure of the LASCO clock in 1998 and the drift of the onboard SOHO clock. We note that the Level-1 data does not remove optical artifacts such as stray light, hot or cold pixels, or cosmic rays. It should be assumed from here on in this article that all references to the LASCO data refers to this Level-1 data product, and that all processing steps incorporate only the nominal full-resolution LASCO C2 observations (1024$\times$1024 pixel resolution, orange filter, clear polarizer). Such observations comprise the overwhelming majority of C2 data.

The construction of the LASCO CBI requires a number of steps be taken to appropriately prepare the LASCO observations. These steps are summarized below and detailed in the following Sections.
\begin{enumerate}[label=\roman*.]
\item Creation of a ``daily median'' of all available daily LASCO C2 observations;
\item Creation of ``monthly minima'' background models from the daily median data;
\item Subtraction of an appropriate background model from each daily LASCO observation and compilation of these data into ``subtracted daily median'' observation;
\item Extraction of the coronal brightness indices via extraction of radial scans from the subtracted daily median data.
\end{enumerate}

\subsubsection{Creation of Daily Median Observations}\label{s:mk_daily_meds}
A key goal in constructing the CBI is to produce a data product that reduces the entire LASCO mission archive into a single data product that can quickly and efficiently be downloaded, stored in computer memory, and analyzed. Furthermore, the CBI is targeted to enable studies of the solar corona over multiple solar cycles.

Thus, the first step in producing the CBI is to produce a single data cube of all appropriate C2 images available during each entire day, and record the median value of each pixel for that day, resulting in a single 1024$\times$1024 ``daily median'' image. Our data selection criteria omitted days in which less than five such observations were available. There also exist a number of days in which no observations were recorded, most notably during 1998 when communications with the SOHO spacecraft were lost between June and September of that year. This process removes the effects of randomly occurring energetic particle and cosmic ray ionization tracks through the image, the effects of solar transient events such as a coronal mass ejections (CME), and stellar and (most) planetary transit signals.

Our complete data set spans 16 April 1996 through 31 July 2017, with 7,094 days of observations within that 7,777-day period -- approximately 91\% coverage. The majority of missing data are within the first three years of the mission. The number of days of data for each calendar year are listed in Table~\ref{t:data days}.

\begin{table}
\caption{ Number of daily median data files created for each year between 10 April 1996 and 31 July 2017.}
\label{t:data days}
\begin{tabular}{cc|cc|cc|cc}     
  \hline                   
Year & No. Days & Year & No. Days & Year & No. Days & Year & No. Days \\
  \hline
1996 & 177 & 2002 & 353 & 2008 & 357 & 2014 & 365 \\
1997 & 312 & 2003 & 338 & 2009 & 355 & 2015 & 365 \\
1998 & 222 & 2004 & 348 & 2010 & 353 & 2016 & 363\\
1999 & 294 & 2005 & 361 & 2011 & 357 & 2017\tabnote{Partial year; end of data range.} & 243\\
2000 & 350 & 2006 & 357 & 2012 & 359 & -- & -- \\
2001 & 348 & 2007 & 353 & 2013 & 365 & -- & -- \\
  \hline
\end{tabular}
\end{table}

\subsubsection{21-day Minima Background Models}

For the CBI to contain signal from only the solar K (electron) corona, it is necessary to produce models of the additional signals present in the observations which can then be subtracted from the daily median product. Specifically, these signals are the F, or dust, corona, stray light, and, to a much lesser extent, very bright planet transits. (The latter are impossible to remove but make negligible impact upon the CBI.) To construct these ``21-day minima background models'' (hereafter referred to simply as ``background models'') we follow the procedures discussed in \cite{Morrill06}, and described as follows.

Every seven days, we obtain a single 1024$\times$1024 image based on minimum pixel value of the surrounding 21 days of daily median data, resulting in 52 so-called background models per year. If a full 21 days is not available in any given period, the maximum number of available days in that period are used, ensuring at least nine daily median data files are available. The background models contain scattered photospheric light due to inner solar system dust, instrumental artifacts such as the occulter pylon, and both continuous and sporadic stray light features.

Production of these background models, and their subsequent subtraction from the daily median data, is a critical step of the processing. Without the step, the resulting indices described later in this article suffer from a number of inconsistencies and discontinuities as a consequence of SOHO's orbit, spacecraft rolls, and changing characteristics of the background stray light pattern (described in an upcoming publication of the LASCO calibration, currently in progress).

\subsubsection{Background Removal Process}

To produce our final data product, we again use the individual level-1 data, this time subtracting an appropriate interpolated background model from each individual image before compiling them into a single background-subtracted daily median file as described in Section~\ref{s:mk_daily_meds}. Specifically, for any given individual data file at some observation time, the two adjacent background models are identified and a linear interpolation made between them to create a background model for the specific observation.

As noted in \cite{Morrill06} the background models typically contain a small residual of the K (electron-scattered) corona, and thus by way of their subtraction we do remove a small amount of K coronal signal from the daily median data, primarily in persistent streamers in equatorial latitudes and most noticeably around periods of low solar activity. Via analyses to be presented in a future publication of the LASCO calibration, we found that excess K corona comprises 1.76\% $\pm$0.40\% of the background models. This value is consistently small throughout the mission timeline, and enables us to conclude that our process for generating background models of the F corona is of sufficient long-term stability that derived indices can be reliably trusted with respect to the quantity of the K corona removed.

\begin{figure*}
  \centering
  \includegraphics[width=110mm]{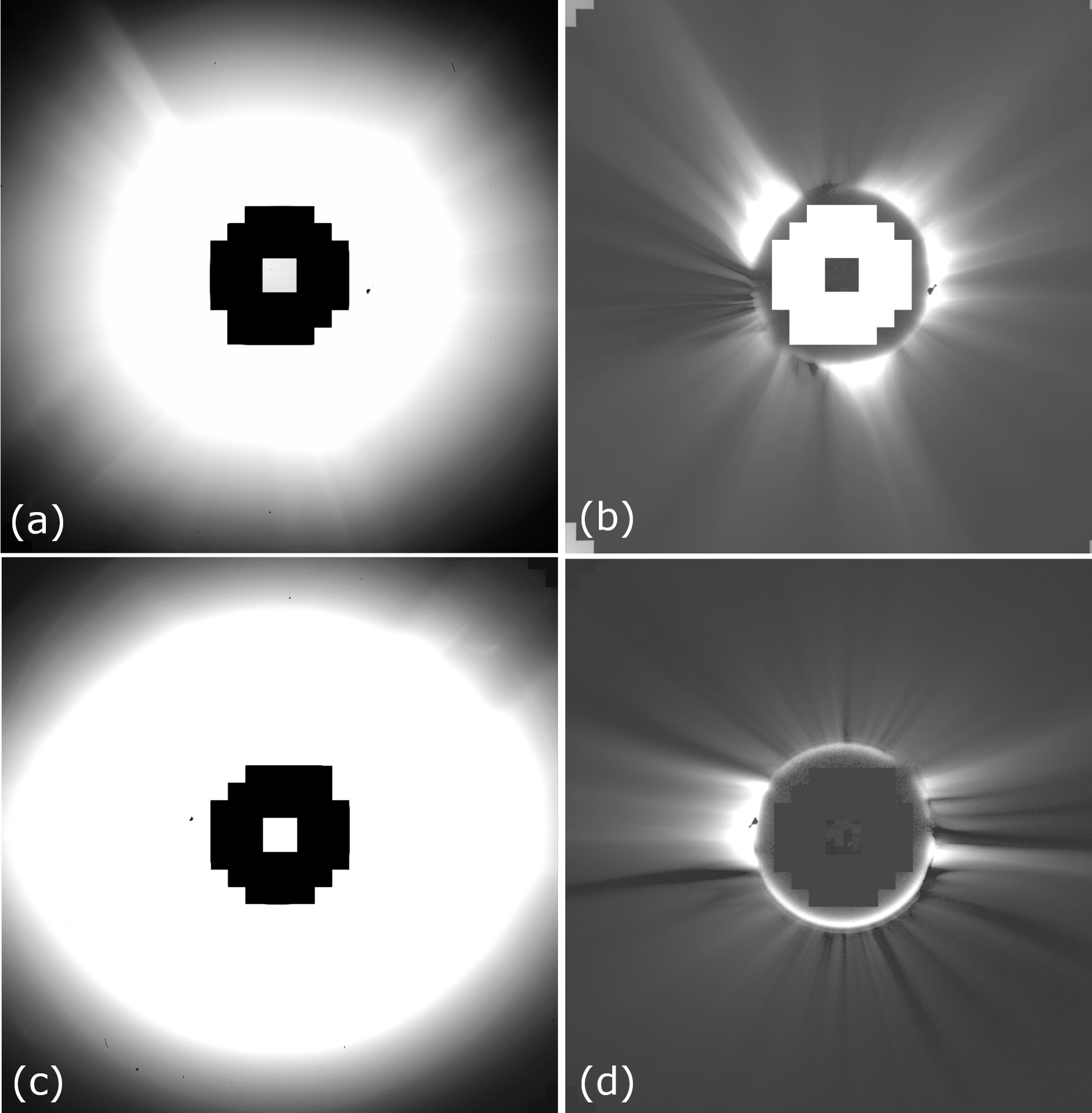}
  \caption[Images]{The LASCO C2 daily median image for Jun-01-2000 before (a) and after (b) the removal of the corresponding ``monthly minimum'' background image, and the same for Jun-01-2009 before (c) and after (d) the background removal. The former date is representative of an observation taken near solar maximum, and the latter near solar minimum. The apparent size of the F-corona `football' in (c) is a consequence of imaging scaling, and should not be interpreted as a structurally different F-corona at solar minimum.}
\label{fig:fcor_removal}
\end{figure*}

The result of the monthly minimum subtraction process can be observed in Figure~\ref{fig:fcor_removal}. The top row shows a daily median image before (Figure~\ref{fig:fcor_removal}a) and after (Figure~\ref{fig:fcor_removal}b) the removal of the monthly minimum image for an image taken near solar maximum on June 01, 2000. The bottom row shows a daily median image before (Figure~\ref{fig:fcor_removal}c) and after (Figure~\ref{fig:fcor_removal}d) background removal for an image taken near solar minimum on June 01, 2009. The apparent size of the F-corona `football' in Figure~\ref{fig:fcor_removal}c is a consequence of imaging scaling, and should not be interpreted as a structurally different F-corona at solar minimum.

\subsection{Construction of the CBI}

Our final data product -- the LASCO CBI -- is constructed by extracting values from the background-subtracted daily median data files at specific locations in the corona. Specifically, for each day of observation, we extract 360-degree radial profiles [${\theta}$] at 38 heights [$r$] in the corona from 2.4{$\mathrm{R}_\odot$} to 6.2{$\mathrm{R}_\odot$}, providing us with a 360$\times$38 element array for each day of observation.

Of critical importance to include in the brightness extraction algorithm are steps that correct for i) the Sun--spacecraft distance, which alters the pixel plate scale values; ii) the variation in the pixel value of Sun-center in the observations, which varies as a function of spacecraft roll; and iii) the spacecraft roll angles, which have varied throughout the mission. Data values are extracted such that all radial profiles begin at a zero-degree position angle (corresponding to solar north) and proceed counter clockwise through the LASCO field of view such that 90-degrees corresponds to the East limb, 180-degrees to the solar south pole, and 270-degrees to the West limb. These steps ensure that our final CBI product is spatially co-aligned, enabling studies of the long-term variability of specific coronal ($r\theta$) locations, in addition to global trends in coronal brightness. What remains is a 360$\times$38$\times$7094 day data cube, and constitutes our final CBI data product.

As noted, missing days of observations naturally lead to discontinuities in the CBI time series. Therefore, to facilitate studies that may wish to, for example, correlate CBI to other geophysical indices (as demonstrated in Section~\ref{s:sci-driven-invest}), we have created an additional CBI product in which we linearly interpolate across all data gaps such that the CBI is a continuous record from the first to the last observation. This corresponds to a CBI data cube of dimensions 360$\times$38$\times$7777 days. Some of these interpolations in the early mission are beyond a length at which they could be considered physically representative of coronal brightness over that period, and thus should be used with caution.

\subsection{Results}

\begin{figure*}
  \centering
  \includegraphics[width=115mm]{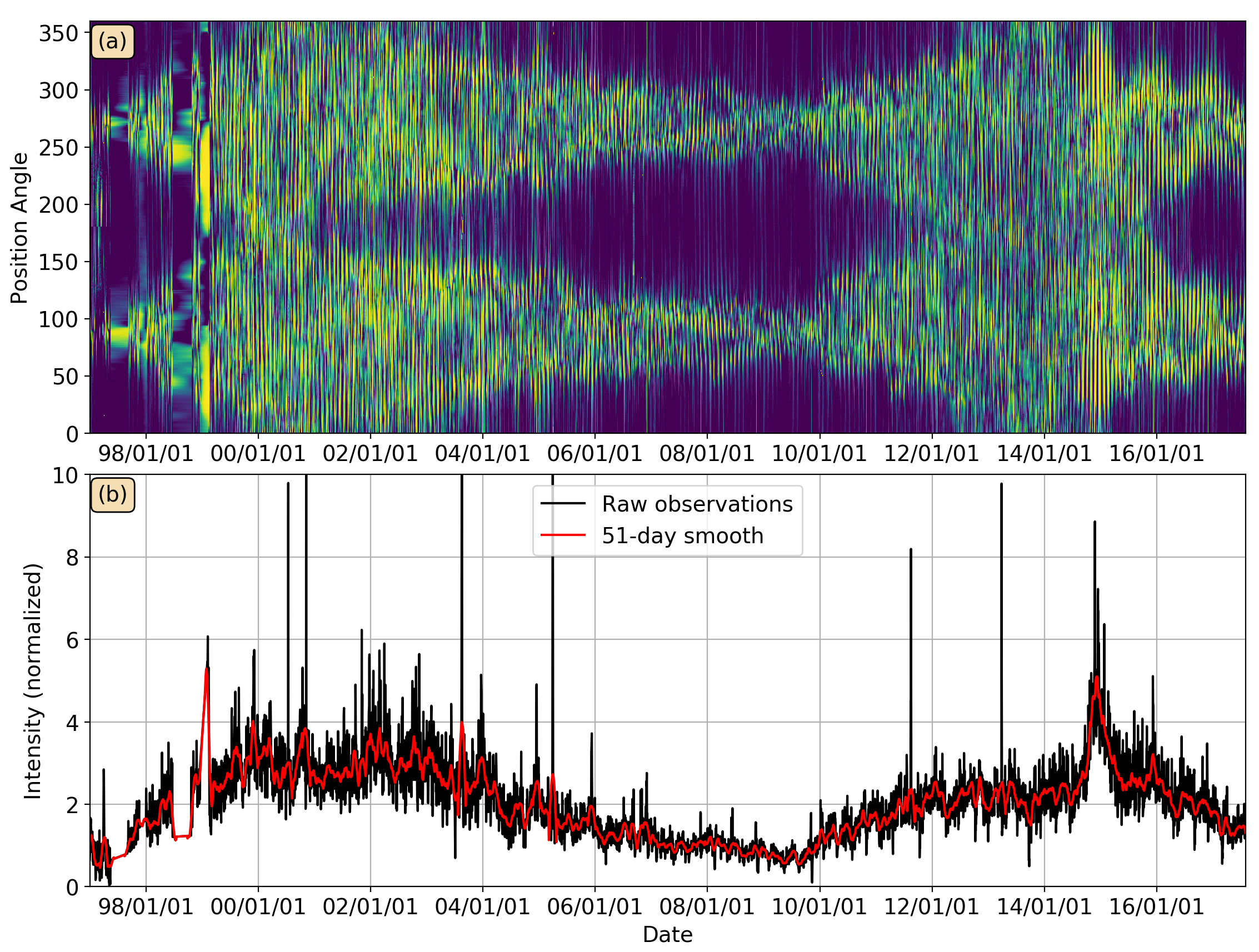}
  \caption[Images]{Visualization of CBI recorded at a height of 5.0{$\mathrm{R}_\odot$} (panel a) from 01 Jan 1997 to 31 Jul 2017. Each thin vertical line represents a 360-degree scan extracted from the daily median images at the given height in the corona. The $x$-axis represents time and the $y$-axis position angle. Intensity values are not stated as the image has been normalized and histogram equalized for better visual presentation. Notable is the increase in brightness and high-latitude presence of coronal streamers around the solar maxima, and corresponding dimming and diminished streamer presence at solar minima. Shown in panel b (black; ``Raw observation'') is the time series of the CBI determined from the sum of the daily observations at 5.0{$\mathrm{R}_\odot$} -- literally a sum over position angle of the observations shown in panel a. Overlaid on this (red; ``51-day smooth'') is a 51-day moving average of the observations. Intensity values have been normalized to a zero-minimum.}
\label{fig:lasco_plots}
\end{figure*}

In Figure~\ref{fig:lasco_plots} we show two different possible representations of the LASCO CBI, noting that many different such representations are possible. The upper panel (Figure~\ref{fig:lasco_plots}a) is a visualization of 360-degree radial scans from the interpolated CBI, located at a height of 5.0{$\mathrm{R}_\odot$} in the corona from 01 Jan 1997 to 31 Jul 2017, corresponding to a 360$\times$7517 array. Data from 1996 are omitted here due to numerous operational data gaps in that year's record. The vertical $y$-axis represents the position angle in the corona, and the horizontal $x$-axis representing time, with one vertical ``stripe'' per day. In this visualization we clearly observe the changing latitudinal extent of corona streamers throughout the two solar cycles.  Differences between the two minima in 1997 and 2007-2009 are also evident, with the latitudinal extent of the white-light coronal brightness being larger during the latter minimum than during the previous minimum. This effect can directly be observed by comparing Figure~\ref{fig:fcor_removal}b (recorded at solar maximum) and Figure~\ref{fig:fcor_removal}d (solar minimum).

In the lower panel, Figure~\ref{fig:lasco_plots}b, the black line (``Raw observations'') present a time series of total coronal brightness at 5.0{$\mathrm{R}_\odot$}, obtained by simply summing over the vertical axis (position angle) of the data shown in Figure~\ref{fig:lasco_plots}a. The intensity values have been normalized to a zero-minimum value for easier interpretation. Here we observe primarily the trends in global corona brightness throughout the solar cycle. Of note, we observe the anomalously massive active region AR12192, visible around October 2014, which dominated the coronal brightness for a period of several weeks and resulted in the strongest (smoothed) global brightness values recorded during the entire LASCO mission archive. Overlaid upon the raw observations is a 51-day smooth (red; ``51-day smooth''), with such smoothed time series -- treated as such to remove effects of solar rotation -- particularly useful in studies seeking to correlation coronal brightness indices to various geophysical indices. CBI time series such as this can be obtained in a number of ways from the CBI data set, as discussed in Section~\ref{s:sci-driven-invest}.

\subsection{Spectral Properties of LASCO CBI}
\label{s:CBI-Spectra-Properties}

In Section~\ref{s:sci-driven-invest} we present science-driven investigations using the CBI in which we demonstrate how the CBI can be used to i) produce a low-resolution proxy for a number of solar irradiance indices and ii) investigate the spatial relationship of the LASCO C2 solar corona and any arbitrary geophysical index. In both examples, and any number of similar studies, we find the CBI data often benefit from smoothing over some number of days (or solar rotations) to filter periodic signals from the data that arise due to a combination of spacecraft operations and solar rotation. In this Section we briefly explore the spectral properties of the LASCO CBI and demonstrate the effect that smoothing (filtering) the CBI timeseries has upon correlation-based studies.

\begin{figure*}
 \centering
 \includegraphics[width=120mm]{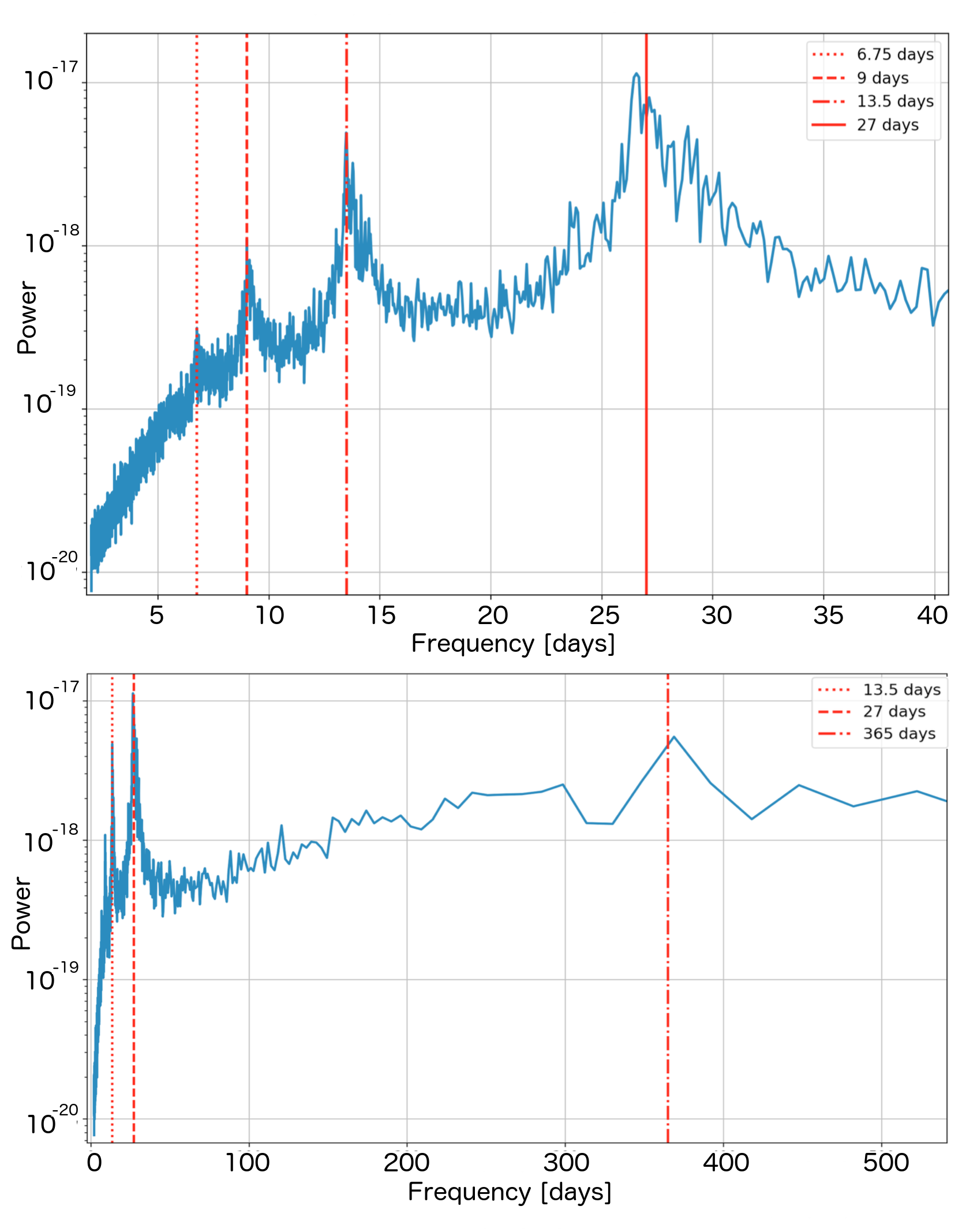}
 \caption{The average power spectrum of all CBI time series. The upper panel focuses on shorter periodicities, and the lower panel on longer frequencies. The power spectrum beyond 500 days is too noisy to make inferences regarding signals in the time series. The vertical lines indicate the periodicity of the dominant peaks, but are not values derived from the observations.}
\label{f:cbi-power-spectrum}
\end{figure*}

Figure~\ref{f:cbi-power-spectrum} presents two portions of the complete average power spectrum of all Spatially Located CBI time series. That is, we computed the power spectrum of each individual time series for each of the 38 radial $\times$ 360 angular element time series in the CBI, and calculated the average power spectrum of all 13,680 time series. We note that Figure~\ref{f:cbi-power-spectrum} only shows the portions of the power spectrum in which we observe significant periodicities.

In the upper panel of Figure~\ref{f:cbi-power-spectrum} we see four dominant peaks in the spectrum at 27, 13.5, 9 and 6.75 days, indicated by the vertical lines in the figure. The peaks observed in this figure are essentially identical to those identified by those shown in Figure 2 of \cite{Parker82}, though that work does not give details on the origins of these signals, instead investigating coronal rotation rates as a function of latitude. The 27-day signal we observe is a consequence of the 27-day solar rotation, indicating that the solar corona retains some degree of similarity over the course of an entire rotation, with the relative breadth of that peak likely a consequence of the $\approx$27-33 day differential rotation rates of the solar surface (which themselves do not directly translate to rotation rates of coronal features). The 13.5-day periodicity represents half a solar rotation and essentially indicates that a given feature on one limb of the Sun will appear on the opposite limb in 13.5 days. The weaker 6.75-day periodicity is most likely a consequence of a quarter solar rotation - specifically the effect of a feature on any given solar limb being visible at 45-degree either side of the plane of the sky. The 9-day signal is possibly an artificial beat periodicity induced by the regular superimposition of 6.75- and 13.5-day signals, \textit{i.e.}

\begin{equation}
f_{\textup{beat}} = (1/f_1 + 1/f_2) / 2 = (1/6.75 + 1/13.5) / 2 = 1/9
\end{equation}

\noindent
particularly as there is no known operational (spacecraft/instrument) or regular physical (coronal) process that produces 9-day periodicities, and a one-third solar rotation signal is not plausible in the absence of a corresponding two-thirds (18-day) signal. \cite{Parker82} does not offer any explanation of the 8.9-day peak observed in Figure 2 of that article. However, 9-day signals have been observed in solar wind streams related to coronal holes \citep{Temmer07,Lei08a} and in Earth's thermospheric densities, theorized as a consequence of solar wind streams from a series of coronal holes spaced approximately 120-degrees apart \citep{Lei08b, Deng11}, though the latter note that ``the 7 and 9 days apparently reflect subharmonics of the 27-day rotation''. Whether any direct connection exists between these reported phenomena and the LASCO CBI is currently undetermined, but warrants further investigation.

In the lower panel of Figure~\ref{f:cbi-power-spectrum} we show an expanded portion of the spectrum that shows an annual peak around 365 days (not exactly at 365 due to the sampling uncertainties), a consequence of the 365-day spacecraft orbit around the Sun.

In Section~\ref{s:irradiance-proxy}, we present and detail a solar irradiance proxy derived from CBI observations smoothed by 81 days (three solar rotations). The choice of 81 days was selected such that we could be confident that all but the most persistent solar features would be removed from the CBI signal. However, shorter smoothing parameters do still provide strong proxies, with the correlation coefficient between the CBI-derived proxy and the chosen irradiance index a function of the level of smoothing applied, as demonstrated by the following analyses.

\begin{figure*}
 \centering
 \includegraphics[width=120mm]{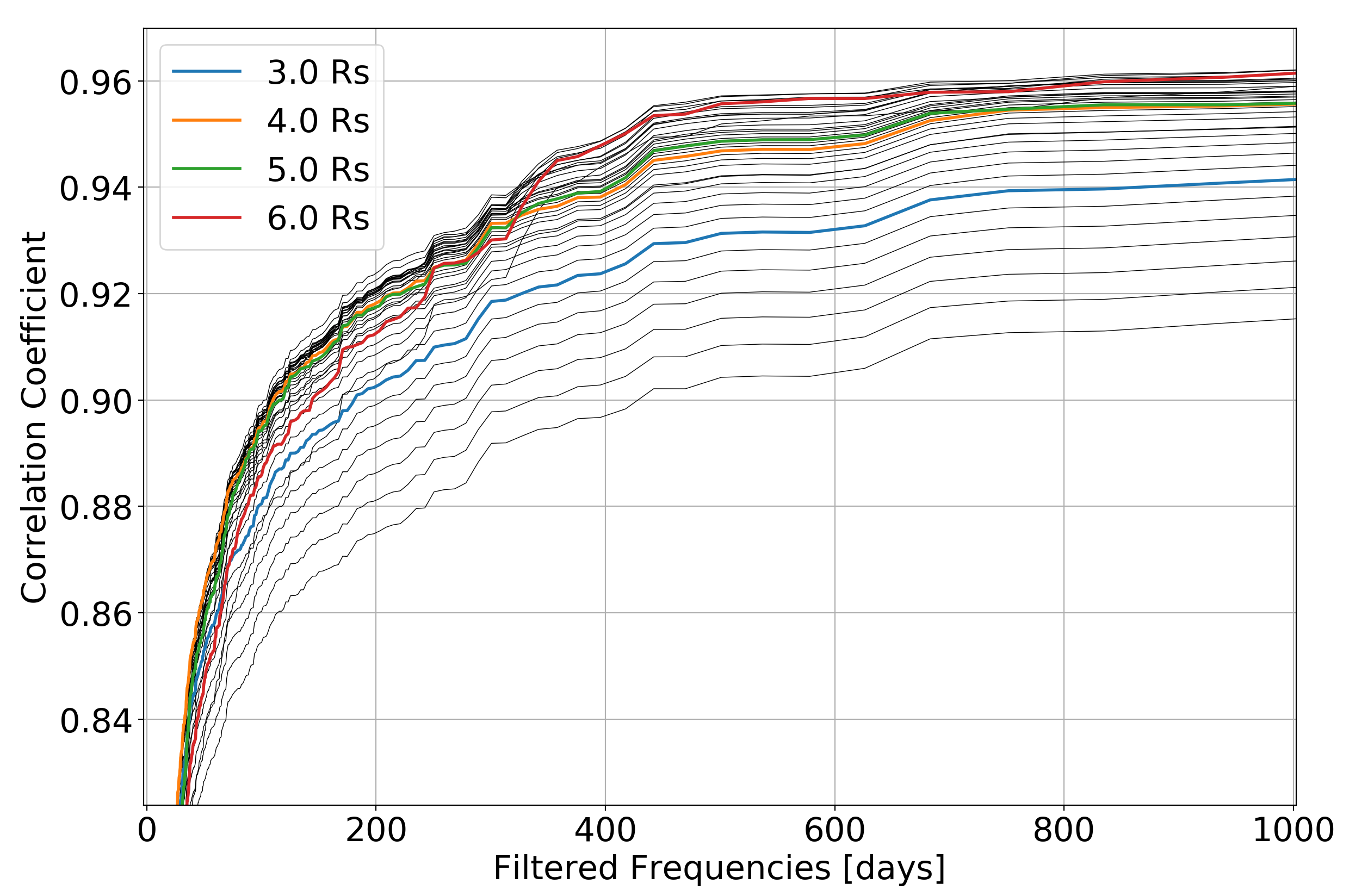}
 \includegraphics[width=120mm]{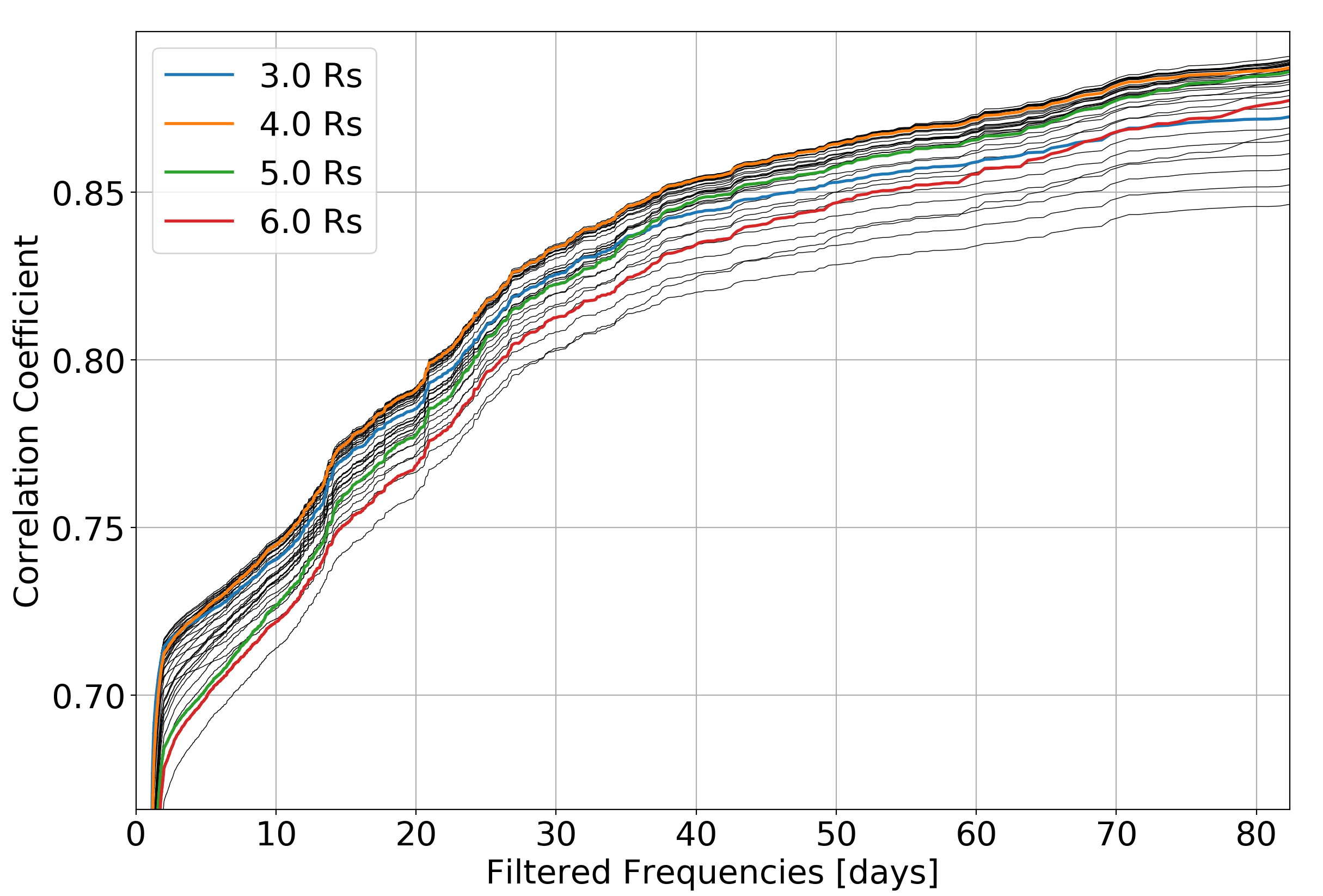}
 \caption{\textit{Upper Panel}: the frequency filtering response of CBI versus NRLTSI2 correlation coefficients for 0--1000 days. This curve extends to the full length of the data set but is essentially flat beyond day 1,000, thus we present here a smaller sample of the large curve. \textit{Lower Panel}: a close view of the curves from the upper panel between frequency values of 0 -- 82 days. In both panels, each black line corresponds to a different height in the corona, with select lines labeled at four discrete radial intervals in the corona.}
\label{f:prog-filt}
\end{figure*}

In Figure~\ref{f:prog-filt} we present the frequency filtering response of CBI versus NRLTSI2 correlation coefficients for the 38 radial CBI indices extracted at each discrete 0.1$\mathrm{R}_\odot$ increment. For each index, we apply an n-day smoothing to both the CBI and the NRLTSI2 irradiance index, and find the correlation coefficient between them. We then increment the n-day filter by one day and repeat. As expected, the correlation coefficients tend towards 1.0 as n approaches the number of days in the time series. Specifically, for Figure~\ref{f:prog-filt} the upper panel shows the frequency filtering response of CBI versus NRLTSI2 correlation coefficients for 0--1000 days. This curve extends to the full length of the data set but is essentially flat beyond day 1,000, thus we present here a smaller sample of the large curve. The lower panel of Figure~\ref{f:prog-filt} shows a close view of this curve for filtered frequency values of 0--82 days. In both panels each black line corresponds to a different height in the corona, with select lines labeled at four discrete radial intervals for reference. These coronal heights are not singled out for any notable reason, and we observe no global trend in coronal height versus correlation coefficient in this data set.

The curves in Figure~\ref{f:prog-filt} show clearly that filtering (smoothing) of just a few days has a significant impact in correlation coefficients, as short-term signals present in the CBI observations, but absent from NRLTSI2, are removed. In Section~\ref{s:sci-driven-invest}, where we present a detailed study of this correlation, we apply an 81-day smooth to the observations to derive CBI-based proxy of a number of irradiance indices. However, as can be seen from Figure~\ref{f:prog-filt}, this smoothing parameter could be reduced to, say, 40 days with only minimal impact on correlations.

Also of note in Figure~\ref{f:prog-filt} are apparent jumps in correlation coefficients at certain filtering values. In the lower panel particularly, we see notably discontinuities in the curves at filtering frequencies of $\approx$13 days and 21 days. These jumps correspond to the removal of frequencies that are present in one - but not both - of the indices being compared. In Figure~\ref{f:cbi-power-spectrum} we observe a clear signal at 13.5 days that is not present in the NRLTSI2 time series (power spectrum not presented here). The 27-day signal is not observed as a jump in Figure~\ref{f:prog-filt} as it is common to both signals, and the 21-day jump seen in Figure~\ref{f:prog-filt} is present only in the NRLTSI2 signal, and hence its removal causes a small but noticeable increase in correlations.

The spectral features we observe in LASCO coronal brightness indices are largely expected and generally consequences of spacecraft operations (orbit) and solar properties (rotation). As we have demonstrated, it is informative to be aware of these signals when analyzing the observations. In particular, when looking to compare LASCO CBI with other indirectly related solar time series, we must be appreciative of the fact that both time series will likely contain both solar cycle and solar rotation influences, and thus we must be sure that any relationship we infer between time series is not simply a consequence of shared signals. For example, many solar observations (\textit{e.g.} TSI) will contain a signal related to solar rotation, but many will not contain the smaller harmonics of that, such as those shown in the upper panel of Figure~\ref{f:cbi-power-spectrum}. While inclusion of these signals may not prove detrimental, attempts to use the CBI to produce relative proxies of indirectly related indices, or study correlations between the two indices, may benefit from first filtering one or both data sets. In either case, it is important that users of the CBI be conscious of the presence of periodic signals inherent to the data.

In the following Section, we present a detailed study of the correlation between the LASCO CBI and a number of total and spectral irradiance indices, deriving LASCO-based proxies of these indices. We then demonstrate how the spatial information contained with the CBI data product can be exploited to provide insight into these correlations, and to investigated relationships between coronal brightness and any arbitrary geophysical index.

\section{Science-driven Investigations Using the LASCO CBI}
\label{s:sci-driven-invest}

The remarkable continuity and stability of the C2 instrument, coupled with its precise photometric calibrations, present the opportunity for the investigation of coronal intensities over an unprecedented time line. We now give two approaches demonstrating how the LASCO CBI can be used to construct and investigate indirect proxies of important geophysical parameters, specifically here TSI. Proxies are frequently used throughout the physical sciences for numerous reasons, including aiding with relative cross-calibrations of instrument data sets, filling in gaps in observational data sets where 
an instrument is temporarily absent, or simply providing indirect observations that are proven to be representative of some other index where the desired data simply do not exist. For example, the Disturbance Storm Time ($D_{\mathrm{st}}$) index (an important metric for geomagnetic storms) can be predicted from a solar wind-based proxy \citep{Temerin02}, cosmogenic radionuclide records can be used to produce proxies for solar activity dating back several millennia \citep{JuWu18}, and numerous solar activity indices can be used as proxies for magnetic activity \citep{Broomhall15}. The LASCO CBI constitutes a potentially valuable data product for the creation of various proxies for use in solar and space weather studies.

\subsection{A New LASCO-based Solar Irradiance Proxy}
\label{s:irradiance-proxy}

The Sun is Earth's primary energy source, providing a total irradiance of 1360.8 Wm$^{-2}$ at the average Sun-Earth distance of 1 Astronomical Unit (AU, \cite{Kopp11}). Changes in solar irradiance may alter Earth's climate, atmosphere and ionosphere through radiative, chemical dynamical processes \citep{Gray10, Lean16}. Composite records of TSI include those of the Physikalisch-Metorologisches Observatorium Davos (PMOD, \cite{Frohlich06, Frohlich09}), Active Cavity Radiometer Irradiance Monitor (ACRIM, \cite{Willson03}, \cite{Scafetta14}) and the Royal Meteorological Institute of Belgium (RMIB, \cite{Mekaoui08}, \cite{Dewitte16}). Each TSI composite record is constructed by cross-calibrating and combining a selected subset of individual irradiance observations, beginning in 1978 for the PMOD and ACRIM records and in 1984 for the RMIB record. Long-term trends and inter-minima levels differ among the three different TSI composite records. In the ACRIM composite, the change in TSI from 1996 to 2009 is -0.28Wm$^{-2}$, larger than in the PMOD composite, whereas there is no change in the RMIB composite. Trends in the three different composites since 2003 differ also from direct TSI observations made by the Total Irradiance Monitor (TIM) on the Solar Radiation and Climate Experiment (SORCE) spacecraft \citep{Kopp11}. By virtue of its state-of-the-art stability of 10 ppm per year, the TIM observations can adjudicate between the PMOD, ACRIM and RMIB composite records after 2003 but not their inter minima differences since TIM observations began only in 2003.

Similarly, in spectral irradiance datasets, the 30.4nm observations made by the Solar EUV Monitor (SEM, \cite{Judge98}) on the Solar and Heliospheric Observatory (SOHO) have a somewhat larger downward trend than is present in observations made by the Solar EUV Experiment (SEE, \cite{Woods05}) on the Thermosphere Ionosphere Mesosphere Energy and Dynamics (TMED) spacecraft since 2001 (when the TIMED was launched). Trends in both differ from new observations made by the EUV Variability Experiment (EVE, \cite{Woods12}) on the Solar Dynamics Observatory (SDO). Nor is the inter minima change in a composite of Lyman-$\alpha$ emission at 121.6nm \citep{Woods00} of a few percent as large as in the SEM observations even though Lyman-$\alpha$ and EUV radiation are both formed in the upper chromosphere-transition region of the Sun's atmosphere.

The lack of independent solar spectral irradiance observations in the visible and near-infrared spectrum precludes validation of observations made by the Solar Irradiance Monitor (SIM, \cite{Harder09}) on SORCE. Whether or not the Sun's disc-integrated visible emission actually varies out of phase with overall solar activity, as \cite{Harder09} suggest, depends on the extent of instrumental trends that \cite{Lean2012} show are present in the SIM dataset. The lack of long-term repeatability of spectral irradiance measurements at most wavelengths inhibits establishing the true character of solar irradiance variability on time scales much longer than the 27-day solar rotation.

Models of solar irradiance variability constructed by quantifying the influences of magnetic features - in the form of dark sunspots and bright faculae - on the disc-integrated radiance closely track the observed total and spectral irradiance changes on time scales of solar rotation modulation (\textit{e.g.} \cite{Lean2005}). The models further reproduce changes observed in TSI during the solar cycle \citep{Kopp11}. However, the models are unable to adjudicate among the observational composite records as to the magnitude of inter minima changes since two different models themselves disagree about the magnitude of these changes. One such model, the Naval Research Laboratory Total Solar Irradiance (NRLTSI2) \citep{Coddington15}, constructed from proxy indicators of the sunspot and facular sources of solar irradiance variability, indicates a TSI change of -0.06Wm$^{-2}$ from 1996 to 2009. In contrast the Spectral And Total Irradiance REconstruction (SATIRE) irradiance variability model \citep{Solanki05,Krivova07},  constructed from solar magnetograms, suggests that the changes were larger, -0.22Wm$^{-2}$ from 1996 to 2009. In the NRLSSI-EUV model \citep{Lean2011} the inter-minima differences from 1996 to 2008-2009 in solar extreme ultraviolet radiation (EUV) is 4\%, smaller than the 15\% change evident in SEM observation.

\subsubsection{Construction of a New Solar Irradiance Proxy}

To quantitatively compare the LASCO CBI [$I_{\mathrm{CBI}}$] with a variety of directly observed and modeled solar irradiance time series [$I_{\mathrm{obs}}$] we first use linear regression to transform the CBI to the scale of each irradiance time series. We do this by constructing a model [$I_{\mathrm{mod}}$] of each irradiance time series as
\begin{equation}
I_{\mathrm{obs}}(t) = I_{\mathrm{mod}}(t) + \epsilon
\end{equation}
\begin{equation}
I_{\mathrm{mod}}(t) = a+b \times I_{\mathrm{CBI}}(t)
\end{equation}

Regressing the CBI against a directly observed irradiance time series determines the amount of variance in that irradiance time series that the independent LASCO observations explains. We quantify this with the correlation coefficient, listed in Table~\ref{t:irrad-metrics}. We then examine the residuals, R\textit{(t)}, of the CBI-derived model and the irradiance time to quantify their differences. In particular, a linear trend line fitted to the residuals
\begin{equation}
R(t) = c + d \times t
\end{equation}
provides a measure of the differences in the long-term slopes (from 1997 to 2017) of the new CBI-based irradiance proxy relative to the irradiance time series. For each irradiance time series and its CBI-based model, the residual time series is determined as
\begin{equation}
R(t) = I_{\mathrm{obs}}(t) - I_{\mathrm{mod}}(t) = I_{\mathrm{obs}}(t) - [a+b \times I_{\mathrm{CBI}}(t)]
\end{equation}

Prior to establishing the model coefficients, a and b, each time series is smoothed over 81 days (equivalent to three solar rotations) to minimize the temporal effects of short-term coronal phenomena and planetary transits in the LASCO field of view and of short-term rotational modulation, primarily from sunspots, on solar irradiance. This smoothing minimizes differences in individual time series arising from different expressions of short-term solar magnetic activity \textit{i.e.} to better express their longer term -- solar cycle time scale -- changes. For this analysis, we use the full (interpolated) CBI data set from 1997-01-01 through 2017-07-31. Again, we omit 1996 observations here as the data are quite sparse, but for completeness include them in the final, shared, CBI data products, per Appendix~\ref{s:data-sharing}.

Table~\ref{t:irrad-metrics} provides statistical metrics of the solar irradiance regression models constructed from the LASCO CBI-based index and the residuals. The metrics for TSI are determined using only data common to all time series so that comparisons of these metrics then quantifies relative differences among the time series. The uncertainties in the model slope coefficients [$b$], and the slope [$d$] of the residuals given in Table~\ref{t:irrad-metrics} are estimated by taking into account the autocorrelation in the residuals \citep{Emmert11}. When the 1$\sigma$ uncertainty in the slope exceeds the slope magnitude itself we conclude that the trend in the CBI agrees (differs insignificantly) with that in the time series.

\begin{landscape}

\begin{table}
\caption{ Relationship of the CBI-derived white-light irradiance proxy with directly observed and modeled solar irradiance changes for 81-day smoothed time series from 01 Jan 1997 to 31 July 2017. The uncertainties given for the model coefficients and the slope of the residuals are 1$\sigma$ statistical values, taking into account autocorrelation in the time series.
}
\label{t:irrad-metrics}
\begin{tabular}{ccccccc r@{.}l c} 
  \hline
Irradiance  & Time series &Correlation with & Model Slope  & Std. Dev. of & Slope of \\
    model &  length (days) &  CBI Proxy & Coeff. (b $\pm$ $\delta$b)  & Residuals (Wm$^{-2}$) & Residuals (Wm$^{-2}$)\\
  \hline
PMOD Total   				& 7,583 & 0.85  & 0.67 $\pm$ 0.05   & 0.209			 & -0.020 $\pm$ 0.010 \\
ACRIM						& 6,149 & 0.85  & 0.83 $\pm$ 0.01   & 0.250			 & -0.035 $\pm$ 0.011 \\
RMIB						& 7,583 & 0.86  & 0.66 $\pm$ 0.01  & 0.195			 & -0.015 $\pm$ 0.010 \\
SATIRE Total 				& 7,583 & 0.85  & 0.65 $\pm$ 0.01   & 0.203			 & -0.021 $\pm$ 0.010 \\
NRLTSI2 Total    			& 7,583 & 0.86  & 0.67 $\pm$ 0.01   & 0.196			 & -0.010 $\pm$ 0.011 \\
SEM 30.4 nm					& 7,527 & 0.77  & 2.59 $\pm$ 0.02   & 1.053x10$^{-3}$ & -0.134x10$^{-3}$ $\pm$ 0.042x10$^{-3}$ \\
NRLSSI-EUV model			& 7,583 & 0.87  & 1.94 $\pm$ 0.01   & 0.545x10$^{-3}$ & -0.041x10$^{-3}$ $\pm$ 0.032x10$^{-3}$\\
HI Lyman-$\alpha$ 121.5 nm	& 7,583 & 0.89  & 1.72 $\pm$ 0.01   & 0.440x10$^{-3}$ & -0.035x10$^{-3}$ $\pm$ 0.024x10$^{-3}$\\
NRLSSI2 540-640 nm			& 7,583 & 0.80  & 0.065 $\pm$ 0.001   & 0.023			  & -0.001 $\pm$ 0.001\\
SIM 540-640 nm				& 5,153 & 0.26 & 0.046 $\pm$ 0.002  & --			 & --\\  \hline
\end{tabular}
\end{table}
\end{landscape}

\subsubsection{Total Solar Irradiance }

\begin{figure*}
  \centering
  \includegraphics[width=120mm]{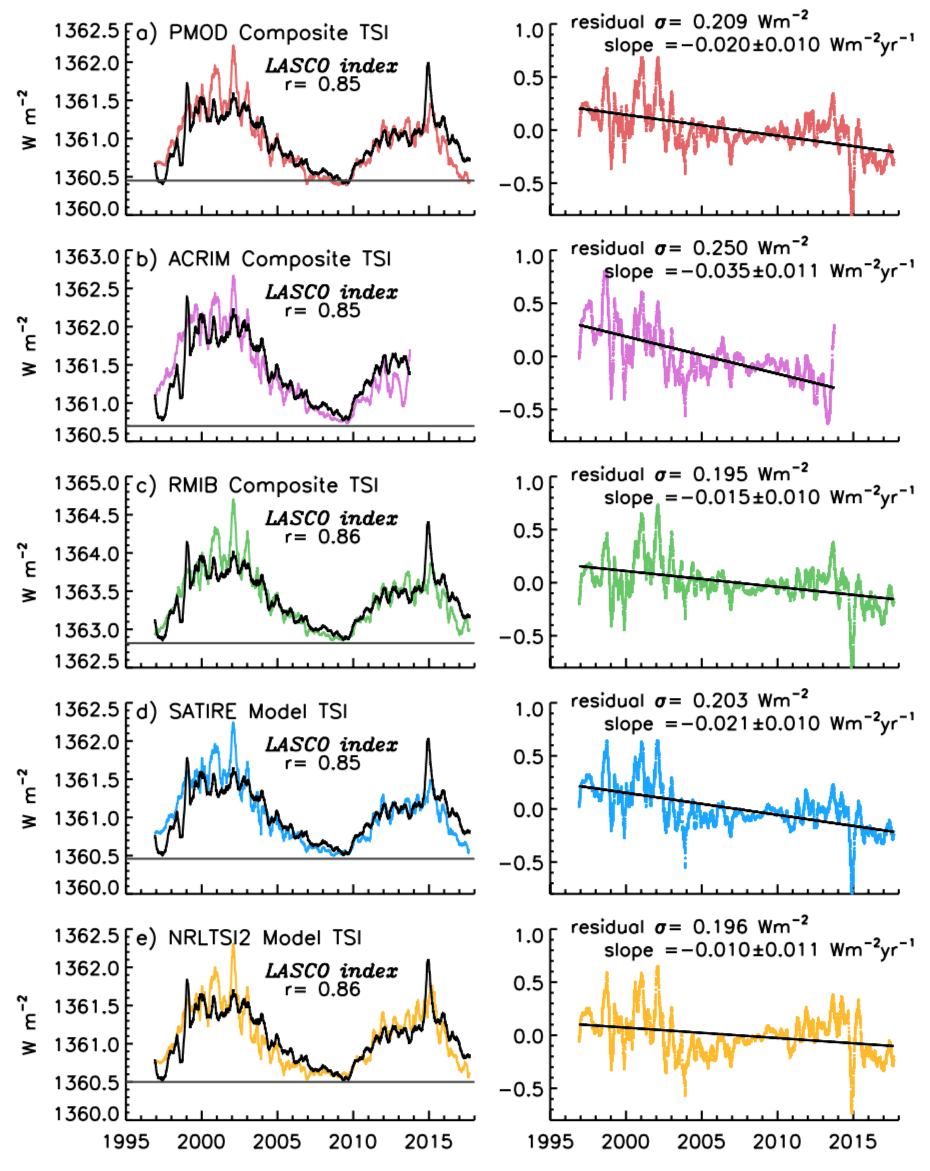}
  \caption[Images]{Comparisons of the LASCO CBI irradiance proxy with five TSI time series are shown on the left, where the LASCO irradiance index time series has been converted by linear regression to a model of the respective TSI time series. Both the irradiance and LASCO time series have been smoothed over 81 days. The correlation coefficient [$r$] for each is indicated.  (a) PMOD composite, b) ACRIM composite, c) RMIB Composite, d) SATIRE model and e) NRLTSI2 model. Shown on the right are the corresponding residuals, determined as irradiance minus LASCO model time series. The 1 $\sigma$ standard deviations and slopes of the residual are also given. }
\label{f:lasco-tsi}
\end{figure*}

Figure~\ref{f:lasco-tsi} compares the TSI variations in the PMOD (Figure~\ref{f:lasco-tsi}a), ACRIM (Figure~\ref{f:lasco-tsi}b) and RMIB (Figure~\ref{f:lasco-tsi}c) composite records with their respective LASCO CBI regression models. Since 1996, the PMOD composite (\textit{Version 42\_66\_1605}) in Figure~\ref{f:lasco-tsi}a comprises observations made by the Variability of solar IRradiance and Gravity Oscillations (VIRGO) instrument on SOHO \citep{Frohlich06}. VIRGO includes two different radiometers, namely that of the Physikalisch-Metorologisches Observatorium Davos (PMOD) and the Royal Meteorological Institute of Belgium (RMIB) DIARAD.  The RMIB composite \citep{Dewitte16} also uses the VIRGO observations, but with different calibration adjustments to the PMOD and DIARD radiatiometers. The ACRIM composite (\textit{Version 131130\_hdr}) in Figure~\ref{f:lasco-tsi} uses observations made by the ACRIM instrument on the Upper Atmosphere Research Satellite (UARS) and ACRIMSAT \citep{Scafetta14}. The CBI irradiance proxy accounts for 86\% of the variance in the PMOD TSI observational composite (correlation coefficient 0.85), 79\% of the variance in the ACRIM composite (correlation coefficient 0.85) and 88\% of the variance in the RMIB composite (correlation coefficient 0.86).

There is good agreement between the trend in the residuals of the NRLTSI2 model (whose inputs are individual sunspot and facular proxies) and the CBI-based proxy as shown in the right-hand panel of Figure~\ref{f:lasco-tsi}e, and also good agreement between the RMIB observational composite and the CBI-based model shown in the right-hand panel of Figure~\ref{f:lasco-tsi}c; the trends in the residuals of 0.015 $\pm$ 0.010Wm$^{-2}$ and 0.015 $\pm$ 0.010Wm$^{-2}$ per year, respectively, are not significant. However, the residuals of the PMOD and ACRIMSAT observations with their CBI proxy regression models, shown in the right-hand panels of Figure~\ref{f:lasco-tsi}a and Figure~\ref{f:lasco-tsi}b, both have slopes that indicate long-term differences.

\subsubsection{Solar Spectral Irradiance }

\begin{figure*}
  \centering
  \includegraphics[width=120mm]{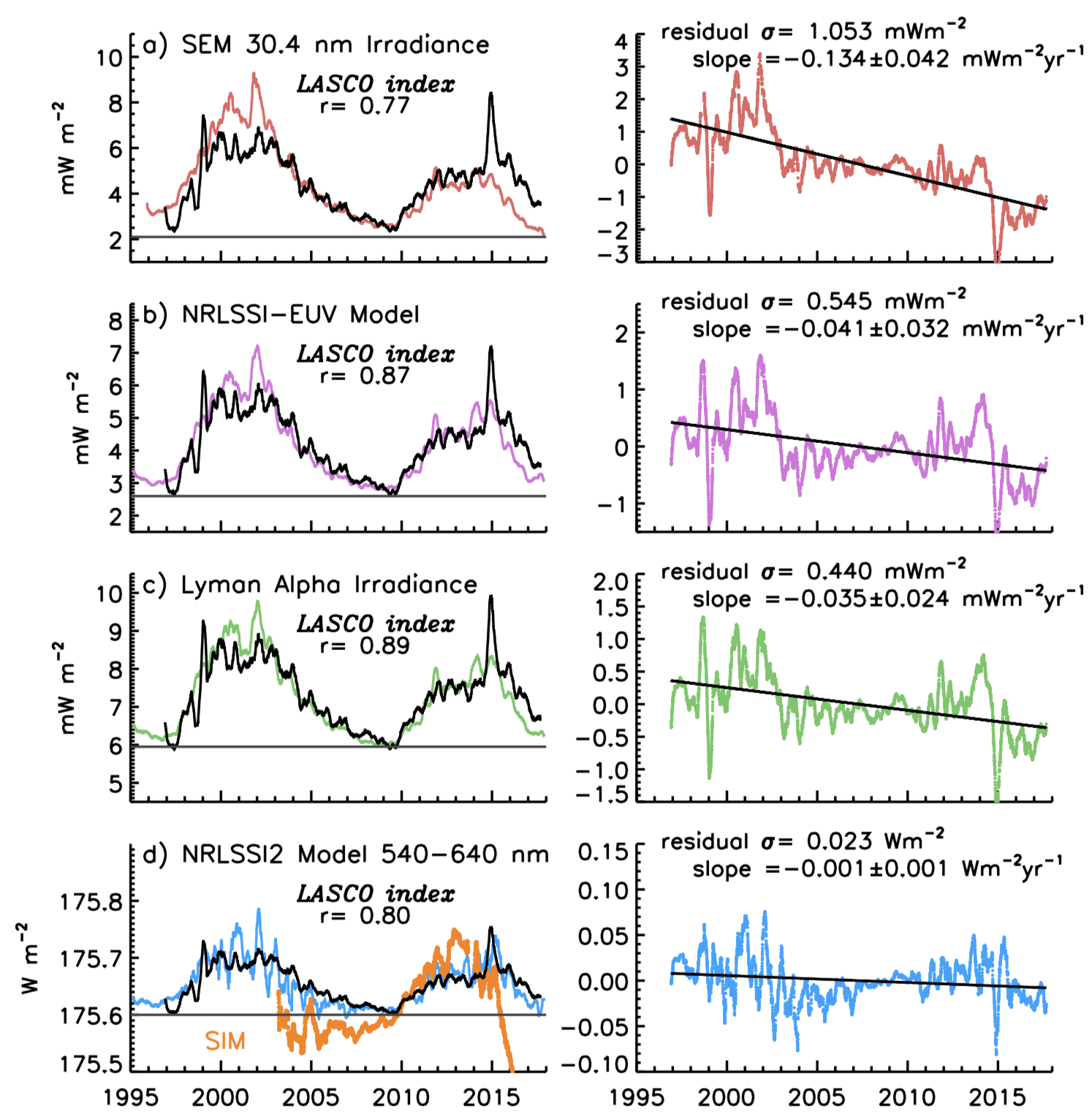}
  \caption[Images]{Comparisons of LASCO CBI proxy with four solar spectral irradiance data sets are shown on the left, where the LASCO irradiance index time series has been converted by linear regression to a model of the respective solar spectral irradiance time series. Both the irradiance and LASCO time series have been smoothed over 81 days. The correlation coefficient [$r$] for each is indicated.  (a) SEM 30.4nm EUV irradiance , b) NRLSSI total EUV irradiance, c) Lyman-$\alpha$ irradiance composite and d) NRLSSI model of 540-640nm irradiance, the wavelength band of the LASCO visible brightness signal, and SIM direct observations.  Shown on the right are the corresponding residuals, determined as irradiance minus LASCO-model time series. The 1 $\sigma$ standard deviations and slopes of the residual are also given.}
\label{f:lasco-ssi}
\end{figure*}

Figure~\ref{f:lasco-ssi} compares the CBI-based irradiance proxy with time series of spectral irradiance. Two of the longest spectral irradiance time series available are those of He II 30.4nm measured by SEM on SOHO and a composite of H {I} Lyman-$\alpha$ irradiance at 121.6nm constructed by the Laboratory for Atmospheric and Space Physics (LASP) from multiple space-based observations; these are shown in Figure~\ref{f:lasco-ssi}a and Figure~\ref{f:lasco-ssi}c respectively. Shown in Figure~\ref{f:lasco-ssi}b, is a model of the integrated EUV irradiance at wavelengths less than 105 nm, estimated by the NRLSSI-EUV proxy model constructed from TIMED SEE observations \citep{Lean2011}. Long-term trends in the NRLSSI2  (Figure~\ref{f:lasco-ssi}d) do not differ significantly from those in the CBI-based proxy, with trends in the residuals of just -0.001 $\pm$ 0.001Wm$^{-2}$ per year. The NRLSSI-EUV and Lyman Alpha data sets (Figure~\ref{f:lasco-ssi}b and Figure~\ref{f:lasco-ssi}c) show a more noticeable long-term trend, with residual trends of -0.0401 $\pm$ 0.032Wm$^{-2}$ and -0.035 $\pm$ 0.024Wm$^{-2}$ per year, respectively. The SEM 30.4nm irradiance data set (Figure~\ref{f:lasco-ssi}a) shows a more significant long-term trend of -0.134 $\pm$ 0.042Wm$^{-2}$ per year, with much of the difference occurring around solar maxima. In general, all Total and Spectral Irradiance data sets show better agreement with a CBI-based proxy during solar minimum, implying that the highly variable nature of the streamer-dominated K-corona during solar maximum is limiting the ability of CBI to describe these irradiance time series.

Observations of the visible and near infrared spectral irradiance commenced only in 2003 with the launch of SIM on SORCE, and these observations are shown in Figure~\ref{f:lasco-ssi}d in the broad visible band of 540-640nm, corresponding to the bandpass of the nominal Orange-filtered LASCO C2 observations of the white-light corona. The SIM visible band irradiance differs notably from the NRLSSI2 model of spectral irradiance variations in this wavelength band, also shown in Figure~\ref{f:lasco-ssi}d. The CBI-based proxy provides a good representation of the NRLSSI2 visible irradiance variability, but is unable to account for any of the variance in the SIM observations (correlation coefficient $0.26$).

\subsection{Spatial Correlation Maps}

Section~\ref{s:irradiance-proxy} demonstrated how a CBI time series can be constructed by summing over all position angles at a given height in the corona to generate a single coronal brightness value for each day of observation. However, one of the unique and valuable aspects of the CBI is the ability to derive many such indices from a number of different coronal locations. For example, studies of equatorial streamers could select angular ``wedges'' at $\approx$90- and 270-degrees covering all coronal heights. Alternatively, indices could be derived from all position angles at some coronal height range (say, 2.5 - 4{$\mathrm{R}_\odot$}), or perhaps within some angular range. The length of the LASCO mission also means that time series can be selected covering certain time periods, for example solar minimum versus solar maximum, or both solar maxima, for example.

The CBI thus presents a valuable flexibility in how time series can be derived and studied. At the very finest resolution, the CBI enables the creation of individual 0.1{$\mathrm{R}_\odot$} $\times$ 1-degree time series - that is, each individual spatial element of the CBI can be considered its own time series, providing a total of 13,680 time series covering the mission duration. An advantage of this fine resolution is that we can employ a ``correlation mapping'' technique, described below, to study the spatial dependence of relationships between the CBI and any arbitrary index of interest.

\subsubsection{NRLTSI2 Correlation Maps}
\label{s:TSI-corrmap}
\begin{figure}
 \centering
 \includegraphics[width=120mm]{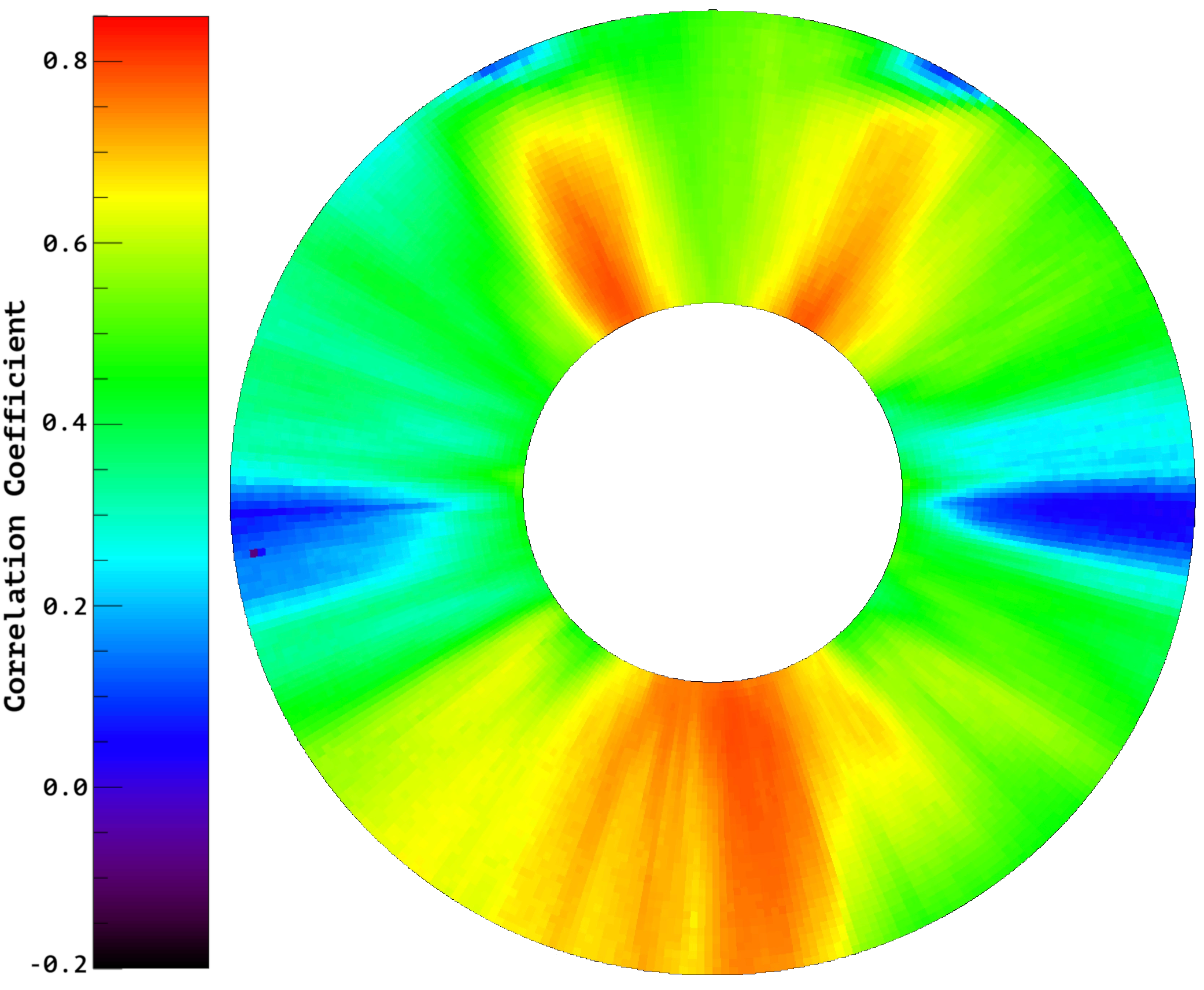}
 \caption{Correlation map for NRLTSI2 versus the LASCO CBI. Here, the NRLTSI2 time series has been correlated to each of the individual 0.1 $\mathrm{R}_\odot$ $\times$ 1-degree spatially located CBI time series, with both data sets smoothed by 54 days (two solar rotations). The statistics for this map are as follows: maximum = 0.85; minimum = 0.10; mean ($\bar{x}$) = 0.61; Standard Deviation ($\sigma$) = 0.14.}
\label{f:cbi-tsi-cmap}
\end{figure}

Following on from our solar and spectral irradiance example, we performed the same steps as in Section~\ref{s:irradiance-proxy} to produce a CBI-derived proxy for the NRLTSI2 index for each $r\theta$ location in the CBI data set, smoothed each time series again by 81 days, and calculated the correlation coefficient between the CBI-based proxy and NRLTSI2 for each coronal location. This process results in 13,680 correlation coefficients that -- for ease of interpretation -- we can map back into the LASCO C2 field of view to produce a ``correlation map'' for NRLTSI2. In Figure~\ref{f:cbi-tsi-cmap} we present such a map, showing the spatial dependence of NRLTSI2 correlations with the LASCO CBI.

From Figure~\ref{f:cbi-tsi-cmap}, we can see that correlations between CBI and the NRLTSI2 model are highly dependent on coronal location, with the polar regions presenting the strongest correlation coefficients up to a maximum 0.85. Conversely, the solar equatorial regions present no meaningful correlation with this index. From Figure~\ref{f:cbi-tsi-cmap} we can propose a plausible scenario in which observations from the polar locations are arguably the least dynamic regions of the corona, and thus present a more ``stable'' electron component over the course of a solar cycle for which incident sunlight to be reflective of the overall energy output of the Sun. A North-South structural asymmetry is visible in this Figure, with two distinct high-correlation regions in the north, and a broader high-correlation region in the south. The source of this structure is somewhat uncertain but likely relates to the average location of coronal structures over the duration of the CBI. An investigation of similar correlation maps for both the North and South components of open magnetic flux \footnote{Obtained from Y.-M. Wang, Priv. Comm.} show remarkably similar structure as shown here, supporting the same result noted in \cite{Barlyaeva15}. A more recent study by \cite{Gomez18} presents their Coronal Density and Temperature (CODET) model that provides clear connections between electron densities and temperatures to (spectral) solar irradiance, showing that the relationship is indeed physical and not one simply of, say, two indices sharing the common 11-year cycle. A detailed investigation of the structure shown in Figure~\ref{f:cbi-tsi-cmap} should yield valuable clues in determining the physical source of the correlation between CBI and solar irradiance.

\subsubsection{Correlation Maps of Other Geophysical Indices}
\label{s:corrmap-other-indices}
\begin{figure}
 \centering
 \includegraphics[width=120mm]{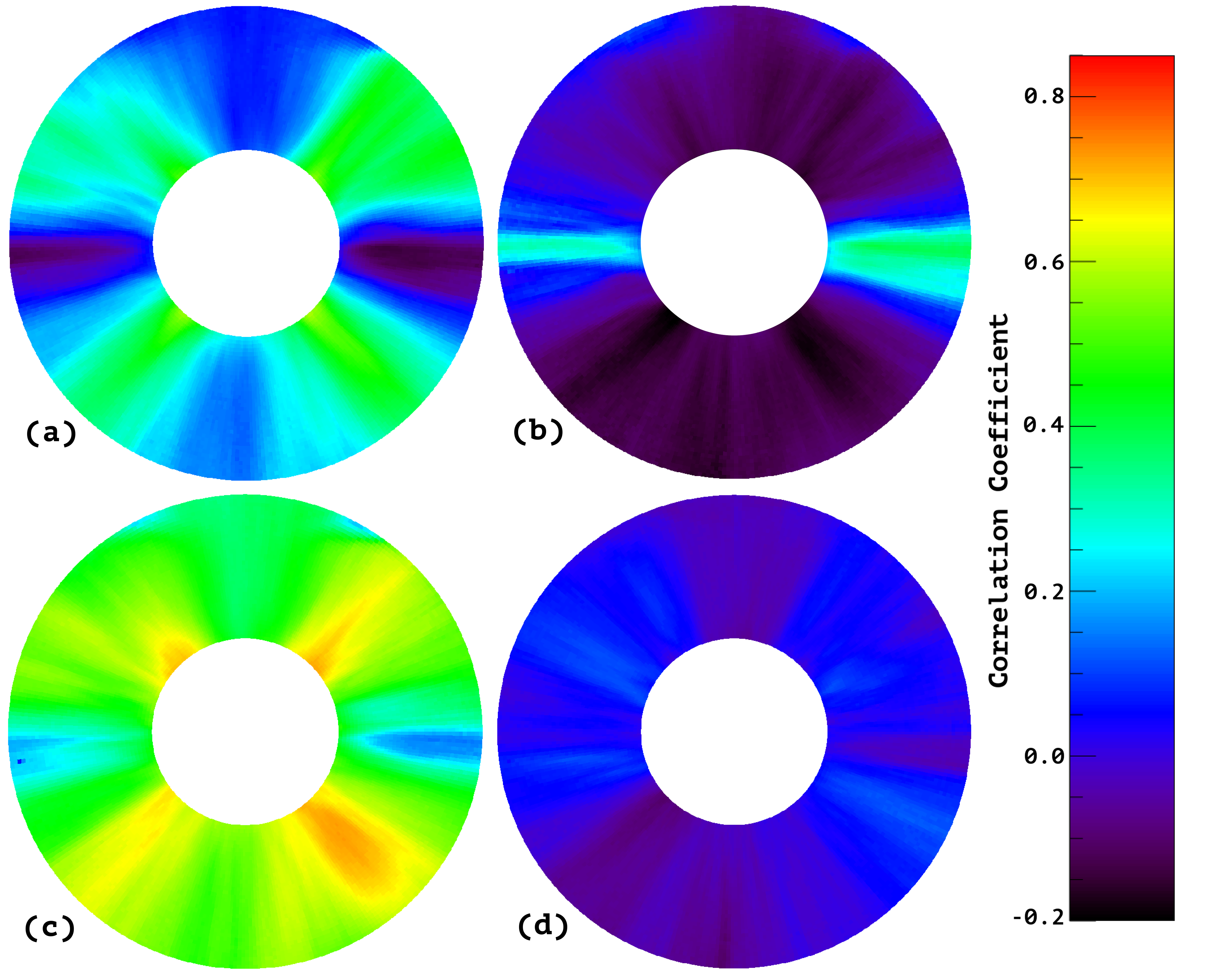}
 \caption{Correlation maps for LASCO CBI versus (a) plasma temperature [K], (b) proton density [n/cc], (c) average interplanetary magnetic field (IMF) magnitude [nT], and (d) the $B_z$ component of the IMF [nT]. The color scales have been adjusted such that all four panels share the same range to enable better by-eye comparison.}
\label{f:other-indices}
\end{figure}

Using the LASCO CBI, it is trivial to extend the methods and analyses presented here to investigate relationships between any arbitrary geophysical index and coronal brightness. In Figure~\ref{f:other-indices} we present a selection of correlation maps comparing the LASCO CBI from 01 Jan 1997 to 31 Jul 2017 to (a) plasma temperature [K] \citep[\textit{e.g.}][]{Hoyt92,Clette16}, (b) proton density [n/cc] \citep[\textit{e.g.}][]{Elliott16}, (c) average interplanetary magnetic field (IMF) magnitude [nT] \citep[\textit{e.g.}][]{Chang73,Boyle97}, and (d) the $B_z$ component of the IMF [nT] \citep[\textit{e.g.}][]{Chang73,Boyle97}. To enable direct comparisons, the color scales have been adjusted such that all four panels share the same range and color table as the data shown in Figure~\ref{f:cbi-tsi-cmap}. These data sets were selected simply as examples of key geomagnetic indices that share a solar origin. They were not selected with any predetermined science goal in mind or in expectation of a strong correlation to coronal brightness, with the $B_z$ data selected specifically because there should be no meaningful correlation between these data and coronal brightness.

All data were obtained from the OMNIWeb service\footnote{See \url{https://omniweb.gsfc.nasa.gov/form/dx1.html}.} and used without modification other than applying a 51-day smooth to remove short-term periodicities. In all cases, the data for these indices covered the entire time period of interest at the required 1-day cadence without data gaps, requiring no interpolation or rebinning for preparation and use. We were thus able to treat each as a single 7,517-element time series that we then correlated against each of the 38 $\times$ 360 individual CBI time series. This produced 38 $\times$ 360 correlation coefficients for each index, which we mapped back into the $r\theta$ domain for visualization (as first demonstrated in Figure~\ref{f:cbi-tsi-cmap}).

In Figure~\ref{f:other-indices}a we observe an apparent weak correlation between the CBI and plasma temperature from 1997--2017, with maximum correlation coefficient of 0.58 ($\bar{x}$ = 0.22, $\sigma$=0.15). The strongest correlations are observed at position angles 45, 135, 225 and 315-degrees, with the poles and equatorial regions showing no meaningful correlation. The observed weak correlation pattern may be a result of CBI correlating to plasma above active regions, which over the course of a solar cycle will tend to populate those regions most commonly. That there is no correlation to equatorial streamer regions implies a fundamental difference between coronal brightness above active regions versus that contained in coronal streamers. This result would benefit from exploration at different stages of the solar cycle. The work of \cite{Gomez18} connecting solar irradiance to coronal electron density and temperature may yield important clues in understanding this relationship.

In Figure~\ref{f:other-indices}b we observe that globally the CBI presents no correlation to proton density, with an exception of a relatively weak correlation in equatorial regions with a maximum correlation coefficient of just 0.4 ($\bar{x}$ = -0.06, $\sigma$=0.13). Here, a possible inference is that the coronal streamers, perhaps mainly those persistent at solar minimum, are presenting a weak but real connection between coronal brightness (CBI) and plasma (proton) density. Again, investigating this at different stages of the solar cycle may yield more information and perhaps improved correlations during certain time periods, though it seems unlikely that any strong relationship should exist between these indices.

Figure~\ref{f:other-indices}c shows a relatively strong correlation between CBI and the average IMF magnitude, with the strongest correlations (maximum = 0.73, $\bar{x}$ = 0.51, $\sigma$=0.10) occurring again in the mid-latitude regions of the corona. The uniformity of the global $\approx$0.5 correlation coefficients may simply be a consequence of both signals correlating to the solar cycle. However, \cite{Yeo17} presents a TSI model based entirely on three-dimensional magnetohydrodynamic simulations of the solar atmosphere, implying that surface magnetism is the primary driver of irradiance variability. Furthermore, some of our investigations of the CBI data product have indicated strong relationships between coronal brightness and magnetic flux in the corona. Thus, a direct connection between coronal brightness (electron content) and IMF magnitude is quite plausible.

Finally, Figure~\ref{f:other-indices}d shows that coronal brightness has absolutely no relations to the $B_z$ component of the IMF. This is an entirely expected result as no plausible connection exists between these two indices. Some spatial structure is observed in the correlation map, but the coefficients are tightly bound between $\pm$0.13 ($\bar{x}$=0.02, $\sigma$=0.05), \textit{i.e.} unequivocally no relationship with CBI. However, this figure panel provides reassurance that our technique can reproduce this known `non-relationship', lending some weight to the more apparent relationships such as that seen in Figure~\ref{f:other-indices}c.

Further detailed investigation/exploration of Figure~\ref{f:other-indices} is beyond the scope of this article, with the purpose of the figure to demonstrate the utility of the CBI and the correlation mapping technique, and demonstrate that other geophysical indices can and do present different spatial correlation structures within the corona \textit{i.e.} that our result shown in Figure~\ref{f:cbi-tsi-cmap} is not solely some artefact of the underlying data processing, coronal structure, or solar cycle.

\section{Discussion}
\label{s:discussion}
LASCO's very long baseline of white light observations of the solar corona, the excellent stability of the LASCO CCD detectors, adjusted by periodic starlight calibrations over the more than 20-year mission lifespan, and the stability of SOHO's L1 orbit, enable the construction of a new index -- the LASCO CBI -- derived from white-light coronal observations with relative photometric precision of about 0.01\%. Whereas previous similar efforts have constructed such indices from radial scans \citep{Dennison12, Lamy14} and equatorial wedges \citep{Barlyaeva15}, we present here -- and provide to the solar community -- a series of fully-calibrated, precisely co-aligned spatial indices obtained at 0.1 $\mathrm{R}_\odot$ $\times$ 1-degree locations in the LASCO C2-observed solar corona from 2.5 - 6.2$\mathrm{R}_\odot$ for the period 14 April 1996 through 31 July 2017. From here on, the CBI will be updated routinely as the instrument PI team produce the final calibrated LASCO level-1 data product.

We demonstrated the utility of the CBI by first producing a proxy of solar irradiance and comparing the proxy to a number of established measures of irradiance. This irradiance proxy (the integrated visible radiation scattered by coronal electrons at 5.0$\mathrm{R}_\odot$) correlates highly with time series of directly observed solar irradiance emitted throughout the solar atmosphere during the solar activity cycle.

The excellent correlation with the NRLTSI2 model gives insight in interpreting the CBI. In that model, the brightness variations of the faculae compete with the darkening due to sunspots. The variation of these two factors are the source of solar cycle variations in irradiance, including in the TSI and the spectral irradiance at most wavelengths. These bright magnetic regions are seen as faculae when observed in photospheric emission and plage when observed in chromospheric emission. That the LASCO coronal data correlates well with solar cycle changes in irradiance suggests that the extension of bright magnetic regions similarly controls solar cycle changes in the extended disc-integrated corona. This interpretation is consistent with prior work that identified correlated brightness variations in solar radiative output from the photosphere to the corona in Solar Cycle 22 \citep{Lean1995}, and consistent also with a recent model that connects spectral solar irradiance to coronal density and temperature \citep{Gomez18}. It is worth reiterating here that the LASCO signal is fundamentally a consequence of photospheric light scattered by coronal electrons along the various lines of sight and is proportional to the product of the two.  The electron density changes on short time scales due to coronal transients (\textit{e.g.} CMEs) and streamer evolution on longer time scales.  On still longer time scales the average electron density changes very slowly, revealing the subtle variability in the photospheric intensity.

Smoothing applied to the CBI time series was observed to increase the correlation, mainly because it minimizes the effects of high frequency fluctuations arising from short-term solar activity that manifests in different ways throughout the solar atmosphere. For TSI, emitted primarily from the photosphere, this smoothing minimizes the effects of short-term irradiance decrease due to sunspots. For the LASCO white-light index, it filters the effects of high frequency coronal fluctuations due to eruptive solar events such as CMEs, thereby exposing the large-scale evolution of the corona. Of course, such treatment of the data means that the LASCO CBI is not intended, nor should be treated, as a high-resolution/high-precision TSI or SSI measurement. Transient events (like CMEs) disturb the short term CBI, and thus reconstructing high-precision TSI or SSI from these data is not possible. However, the longevity and photometric stability of the LASCO observations are such that the CBI can be useful for the long-term study of a number of solar indices.

The ability of the CBI-based irradiance proxy to represent the solar cycle evolution of a variety of total and spectral irradiance observations might be considered somewhat surprising due to the well-recognized large inherent fluctuations in coronal brightness from day to day and over the solar cycle owing to the large variation in the total electron content in the corona. However, the CCD cameras on LASCO have very stable photometry with a relative accuracy throughout the mission on the order of 1 part in 10,000. The LASCO instrument is extremely well calibrated using thousands of stellar transits over the approximately 18 years of spaceflight, enabling the long-term degradation to be well established at 0.2\% per year \citep{Colaninno}, with this degradation principally due to a browning of the optics caused by radiation damage. However, this degradation is corrected during the calibration of the level-1 data product. Thus the CBI irradiance proxy provides a high fidelity measure of relative irradiance variations on the time scales of a solar cycle time over the lifetime of LASCO. That the CBI proxy is able to discern differences in long-term trends among various total and spectral irradiance observations and models shows the sensitivity of the technique.

A key step in preparing the LASCO data is to remove the effects of F-corona wobble, stray light, cosmic rays, and other imaging artifacts, to leave an image comprised almost exclusively of K-coronal (electron-scattered) signal. This is achieved via established background-removal procedures, as described earlier in the text. The LASCO background models do indeed contain a small residual K-coronal signal \citep{Morrill06}, and thus some excess is removed from the data from which we derive our proxy. However, this is largely confined to regions where there is a persistent streamer for a month at the same latitude, which can occur at equatorial regions at solar minimum. In a separate investigation, we found the excess K corona present in the background models (and thus absent from the final CBI) comprises 1.76\%$\pm$0.40\% of the background model and is sufficiently small that most results should be unaffected.

The CBI-derived irradiance proxy represents solar irradiance variations better during the declining and rising phases of the solar cycle and the minima than during solar maximum. We ascribe this to the effects of CMEs passing through the corona and disturbing and temporarily carrying away the large-scale coronal patterns. As many as six CMEs per day are typically observed during the solar maximum period, whereas the rate is about one CME every other day during minimum. As seen in Figure~\ref{f:lasco-tsi}, there is a strong deviation between CBI and the irradiance indices in late 2014. This is due to the anomalously bright solar corona above active region 12192, the largest active region in 25 years. While this active region clearly contributed to a short increase in irradiance, the response of the electron corona was exaggerated significantly. We also note more generally that the residuals of all models (Figures~\ref{f:lasco-tsi} and \ref{f:lasco-ssi}) show a small but noticeable decline over the LASCO mission timeline, though some of these trends (most notably RMIB and NRLTSI2) are not statistically significant. Nonetheless, the origin of this general trend across most indices is unclear and remains under investigation.

Although the CBI-derived irradiance proxy presented in Section~\ref{s:sci-driven-invest} is constructed from observations of white light scattered by electrons at 5.0$\mathrm{R}_\odot$, the correlation with directly measured irradiance is largely independent of distance from the solar limb. Instead, as demonstrated by our correlation mapping technique in Section~\ref{s:TSI-corrmap} that the correlations with irradiance is strongly location-dependent, with high-latitude/polar regions being the primary source of the strongest correlations. While the underlying physical driver of this relationship is somewhat unclear, preliminary investigations imply that magnetic flux in the corona is key in interpreting this result, and that the relative stability (lack of dynamics) in these high-latitude regions are responsible for producing the most stable baseline of magnetic flux (and thus irradiance) proxy observations.

In Section~\ref{s:corrmap-other-indices} we presented correlation maps for four other geophysical indices -- plasma temperature, proton density, average IMF magnitude, and the $B_z$ component of the IMF -- with a goal of demonstrating the broader utility of the CBI. While none of these indices showed relationships to CBI as strong as that of irradiance, we did observe a relatively strong correlation with the averaged IMF magnitude, supporting our earlier conclusions regarding the relationship between coronal brightness and magnetic flux in the corona. We did also observe unique structural relationships between plasma temperature, which correlated strongest in regions above which active regions generally occupy, and proton density, which correlated (weakly) in only the solar equatorial regions. The latter result in particular makes some intuitive sense as plasma contained in solar equatorial regions is most likely that which will also reach Earth.  Finally, we also observe that the $B_z$ component of the IMF shows absolutely no relationship to coronal brightness, as would be entirely expected. As a whole, the correlation maps shown in Figure~\ref{f:other-indices} demonstrate the capability of the CBI to capture not only plausibly real relationships between coronal brightness and arbitrary geophysical indices, but also to enable investigations of the physical mechanisms underpinning those relationships via exploration of the spatial structure of the correlation.

As noted in Appendix~\ref{s:data-sharing}, the LASCO C2 CBI is now available on the LASCO project website, with data files in formats suitable for both IDL and Python environments. As the Level-1 data are periodically processed at the host institution, we will update the CBI data sets accordingly. We plan to explore extending our study to the C3 coronagraph, but note that the larger pixels, the more obvious occulter arm pylon, and fainter coronal structure, will place severe limitations on the data and perhaps render it less suitable to the treatment discussed here. We also plan to investigate applying the technique to coronagraph data recorded by the NASA \textit{Solar Terrestrial Relations Observatory} \citep[STEREO,][]{Kaiser}, where total brightness and polarized brightness images have been recorded daily since January 2007 from two nearly identical spacecraft until Oct 1, 2014, when contact with STEREO-B failed. However, STEREO-A is continuing nominally. These two or three viewpoints should enable unique cross-calibration studies.

\section{Summary}

Utilizing over 20 years of observations of visible radiation emitted by the Sun's surface and scattered by electrons in the solar corona, made by the C2 coronagraph within the SOHO/LASCO instrument, we have constructed a new index (CBI) which reduces the entire mission archive into a single, fully calibrated data set representing the brightness of the solar corona over all position angles and heights from 2.5 - 6.2 $\mathrm{R}_\odot$. By sharing this data with the solar community, we make available a synopsis of the entire LASCO C2 archive, fully calibrated and placed in a single, convenient data file, reducing a significant amount of overhead for any studies that wish to examine the broad, long-term nature of the solar corona.

By virtue of its long duration, covering two successive solar minima in 1997 and 2007-2009, and its high long-term photometric stability, achieved through in-flight monitoring of stars, we are able to utilize the CBI to investigate inter-minima differences in the trends in a variety of solar irradiance datasets during the solar cycle. Statistical comparisons indicate that the recent PMOD TSI observations and the NRLTSI2 model of TSI variability are consistent with the CBI-based irradiance proxy. In contrast, the ACRIM observations and the SATIRE model both have lower TSI levels in the 2008-2009 minima, relative to the 1997 minimum, compared with the LASCO CBI. Based on these comparisons with PMOD and NRLTSI2, we suggest that anomalous long-term trends of -0.029Wm$^{-2}$ and -0.015Wm$^{−2}$ per year (21 and 11 ppm per year), respectively, may be present in these latter TSI time series spanning 1997 to 2014.

The CBI-based irradiance proxy is consistent with inter minima changes in the LASP Lyman-${\alpha}$ composite and the NRLSSI-EUV irradiance model of EUV irradiance variations, but there may be a spurious downward trend in the SOHO SEM observations of 30.4nm irradiance. We find that the CBI-based irradiance proxy of visible radiation in the band from 540 to 640nm scattered by coronal electrons correlates well with this visible irradiance band as represented by the NRLSSI2 model, but has essentially no relationship to the direct SIM observations at 540-640nm. This supports prior evidence of that total and visible solar irradiance both vary in-phase with solar activity \citep{Lean2012} and further suggests the likelihood of long-term instrumental drifts in the SIM observations. With the exception of SIM, we find that the LASCO CBI correlated well with all irradiance indices examined, with coefficients in the range 0.77 -- 0.89.

Via presentation of a new correlation mapping technique, have shown how the structural relationship of the correlation between the CBI and solar irradiance at different coronal locations can yield valuable clues regarding the relationship between the indices, finding that CBI correlation with solar irradiance is located primarily in polar regions and indicating that the relatively stability of the corona in these regions gives rise to the most stable time series for constructing a reliable irradiance index. We then extended this technique to present correlation maps for CBI versus four other geophysical indices, finding that coronal brightness appears to relate quite well to the IMF magnitude and somewhat well to plasma temperature. We also find a weak but plausible connection between proton density and coronal brightness, with this result in particular showing a strong solar equatorial dependency. Finally, we show that no relationship exists between coronal brightness and the $B_z$ component of the IMF, as would be expected from these two physically unconnected observations. The correlation mapping results reinforce our conclusion that studying the unique structural relationships between these indices is a valuable aid in interpreting any possible relationships.

The LASCO observations continue to the present and may aid in preserving the fidelity of various extant geophysical indices, including solar irradiance records, in the event of gaps. The CBI is also a valuable data source for studies seeking to explore the properties of coronal and streamer brightness over the entire mission. We note for example that the massive active region AR 12192, seen around October 2014, not only dominated coronal brightness during that period but is clearly the brightest single coronal source observed over the entire LASCO mission archive. As detailed in the Appendix, we plan to maintain the CBI as a data product made freely available to the heliophysics community, providing data files suitable for both the IDL and Python analysis environments.

%
\acknowledgments
This work was supported by NASA Heliophysics grants to SOHO/LASCO and STEREO/SECCHI, NASA’s Earth Science Solar Irradiance Science Team (SIST), and the NRL Edison Memorial Program. The authors wish to thank an anonymous referee, whose feedback has substantially improved our presentation of these results.

\section*{Disclosure of Potential Conflicts of Interest}
The authors declare that they have no conflicts of interest.

\appendix

\subsection{Data Sharing: Obtaining the LASCO CBI}
\label{s:data-sharing}
We are making available the LASCO CBI as both raw and interpolated data sets, along with the relevant time information, in a number of formats. The primary CBI data product is a 360$\times$38$\times$7094 data cube representing values obtained from existing data files only (\textit{i.e.}, the time series contains many discontinuities). An alternative data cube we provide is a 360$\times$38$\times$7777 data cube representing interpolated observations (\textit{i.e.} linear interpolation across all missing dates). The CBI data cubes are available as both \textsf{IDL} ``.sav'' save files and \textsf{Python/NumPy} ``.npy'' files. File sizes are of the order 406 megabytes for the interpolated data set and 370 megabytes for the raw (not interpolated) data.

Additionally, we provide both ``raw'' and ``full'' data files containing the relevant date information (YYYY-M-D) in the following formats: i) \textsf{IDL} `.sav' files; ii) plain text files; iii) \textsf{NumPy} .npy files containing \textsf{datetime} objects for the dates. An accompanying ``\textsf{README}'' file will contain this information and identify which data products correspond to which filenames. This \textsf{README} file will be updated as the CBI is updated to include new observations. We intend to support and maintain this product through the LASCO mission lifetime.

All aforementioned files are available at the following url: \url{https://lasco-www.nrl.navy.mil/CBI}. Should investigators require any additional metadata, we encourage them to contact the corresponding author of this article. We also request that any publications utilizing the LASCO CBI data cite this publication.

\bibliographystyle{spr-mp-sola}
\bibliography{coronal_brightness_index}

\end{article}
\end{document}